\documentclass{article}
\usepackage[utf8]{inputenc}
\usepackage{amsmath}
\usepackage{amsfonts}
\usepackage{graphicx}

\usepackage{siunitx}
\usepackage{url}
\usepackage{authblk}

\newcommand{\regTag}[1]{\hspace{0.5cm} \textbf{\textrm{(#1)}}}
\newcommand{\revmod}[1]{{#1}}

\begin{document}

\title{Joint Reconstruction in Low Dose \\Multi-Energy CT}
\author[*]{Jussi Toivanen}
\author[**]{Alexander Meaney}
\author[**]{Samuli Siltanen}
\author[*]{Ville Kolehmainen}
\affil[*]{Department of Applied Physics, University of Eastern Finland, POB 1627, FI-70211 Kuopio, Finland}
\affil[**]{Department of Mathematics and Statistics, University of Helsinki, Finland}
\maketitle

% ===================================================================
\section*{Abstract}
Multi-energy CT takes advantage of the non-linearly varying attenuation properties of elemental media with respect to energy, enabling more precise material identification than single-energy CT. The increased precision comes with the cost of a higher radiation dose. A straightforward way to lower the dose is to reduce the number of projections per energy, but this makes tomographic reconstruction more ill-posed. In this paper, we propose how this problem can be overcome with a combination of a regularization method that promotes structural similarity between images at different energies and a suitably selected low-dose data acquisition protocol using non-overlapping projections. The performance of various joint regularization models is assessed with both simulated and experimental data, using the novel low-dose data acquisition protocol. Three of the models are well-established, namely the joint total variation, the linear parallel level sets and the spectral smoothness promoting regularization models. Furthermore, one new joint regularization model is introduced for multi-energy CT: a regularization based on the structure function from the structural similarity index. The findings show that joint regularization outperforms individual channel-by-channel reconstruction. Furthermore, the proposed combination of joint reconstruction and non-overlapping projection geometry enables significant reduction of radiation dose.

% ===================================================================
\section{Introduction}

Computed tomography (CT) is an imaging modality in which the interior structure of an object is reconstructed from X-ray transmission images recorded along different directions. The reconstructed object is typically presented as cross-sectional images or 3D volumetric data. CT imaging plays an ever-diversifying role in medical imaging and industrial applications, with continuing methodological developments being motivated by the desire for increased image quality and information, and for reductions in radiation dose \cite{lell2015evolution, dechiffre2014industrial}. Advances in CT imaging have been driven both by hardware improvements, such as better detectors and X-ray tubes, and by algorithmic developments in reconstruction methods \cite{mccollough2012dose, ginat2014technology}.

A CT image is a greyscale image where the value of a pixel or voxel in a given location is proportional to the X-ray attenuation coefficient of the medium in the corresponding location. A significant limitation of conventional CT imaging is that no bijective relationship exists between the reconstructed greyscale value and the material properties, i.e., elemental composition and mass density. Therefore, two tissues of entirely differing properties may be reconstructed as the same value, a typical example being bone and iodine contrast agent \cite{mccollough2015dual}. More information on the material composition can be obtained by using dual- or multi-energy CT imaging, where the object is imaged using more than one X-ray energy, and the non-linearly varying attenuation properties of different elemental media are exploited for material identification \cite{lell2015evolution,mccollough2015dual}. Applications of multi-energy CT, also known as spectral CT, include differentiation and quantification of materials, tissue characterization, virtual monoenergetic imaging, automated bone removal in CT angiography, cardiovascular imaging, multiple contrast agent imaging, and mapping of effective atomic number \cite{mccollough2015dual, goo2017dual, fornaro2011multi, forghani2018characterization, marin2014characterization, nicolaou2012characterization, wong2018musculosceletal, yu2012mono, postma2012bone, danad2015cardiac, kalisz2017cardiac, desantis2018cardiac,symons2017contrast}.

Multi-energy CT imaging was first proposed in the 1970s \cite{alvarez1976multienergy}, and today there exist many data acquisition schemes for spectral X-ray imaging: sequential scanning, rapid tube voltage switching, dual source scanning, multilayer detectors, and photon-counting detectors \cite{mccollough2015dual}. 
Compared to traditional radiography, CT is a high-dose modality \cite{kalender2014dose}. Although the risks of radiation doses in the range of a few millisieverts remains a controversial topic \cite{tubiana2009radbio,hendee2012risks}, an underlying philosophy in CT development is to reduce the dose as much as possible, in accordance with the ALARA (As Low As Reasonably Achievable) principle \cite{icrp103,icrp105}. There is a concerted effort to push the radiation dose from routine CT examinations to less than 1 millisievert, which is well below the annual dose due to background radiation \cite{mccollough2012dose}. Both image quality and radiation dose depend fundamentally on the number of photons used in imaging, and an acceptable balance needs to be found between the two factors \cite{buzug2008ct}. Radiation dose can be reduced by reducing the number of photons used, and one efficient way of achieving this is to reduce the number of X-ray projections \cite{mccollough2012dose,pan2009fbp,hamalainen2013sparse,sechopoulos2009optimization}. However, image reconstruction from sparse projections requires algorithms more advanced than the standard approach where attenuation at each measured energy is reconstructed independently using classical techniques, such as the filtered backprojection (FBP) algorithm. 

A highly feasible approach for low dose multi-energy CT is to utilize a multi-channel joint reconstruction approach where all the unknown images are reconstructed simultaneously by solving one combined inverse problem. The central idea is to combine all the projection data into a single image reconstruction problem and to utilize regularization models that promote some prior information on the unknown images within and across the energies.
In multi-energy CT, a reasonable prior assumption for the attenuation images at different energies is that they can be expected to be {\em structurally} similar in the sense that an edge (e.g. an organ boundary) that is present at one energy, is likely to be at same location and alignment with the other energies as well, even though the contrast between materials will be different at each energy. There are many models in the literature designed for promoting structural similarity \revmod{of images. These models can work either in a one-sided way, where the model is used to promote structural similarity of the unknown image with a fixed reference image (see e.g. \cite{pan2014structural, yang2012structural}), or two-sided way where the model promotes structural similarity between two (or more) images in a joint reconstruction formulation.} These include vectorial total variation \cite{rigie2015vtv, rigie2017vtv}, spectral patch-based penalty for the maximum likelihood method \cite{kim2015patch}, 
% extension of the classic ART method \cite{zhao2015eart}, 
tensor-based dictionary learning \cite{zhang2017dictionary,wu2018dictionary}, parallel level sets \cite{kazantsev2018joint}, the prior rank, intensity and sparsity model (PRISM) \cite{gao2011prism, yang2017prism}, tensor-based nuclear norm regularization \cite{semerci2014tensor}, nonlocal low-rank and sparse matrix decomposition \cite{niu2018nonlocal}, and total variation regularization using non-convex optimization \cite{chen2017nonconvex}.
%\VilleComment{Jussi: Add here citations to Pan and Yang (ref 1 comment).}

In this paper, we study the performance of various structural similarity promoting joint reconstruction models using a novel low dose data acquisition protocol. We consider 
\revmod{three established joint reconstruction models: joint total variation which was originally developed for denoising of multichannel images \cite{blomgren1998color}, linear parallel level sets \cite{ehrhardt2014vector} and a smoothness regularization similar to the models that have been employed to dynamic MRI in \cite{rasch2018dynamic} and fluoresence x-ray tomography in \cite{gursoy2015hyperspectral}.  
% promoting continuity of the attenuation coefficient in the energy dimension 
Furthermore, we introduce a new joint regularization model 
%for multi-energy CT. One is a spectral smoothness regularization  promoting continuity of the attenuation coefficient in the energy dimension, and the other is a joint regularization model 
based on the structural similarity index \cite{wang2004image}.}

\begin{figure}
\centerline{
\includegraphics[width=6cm]{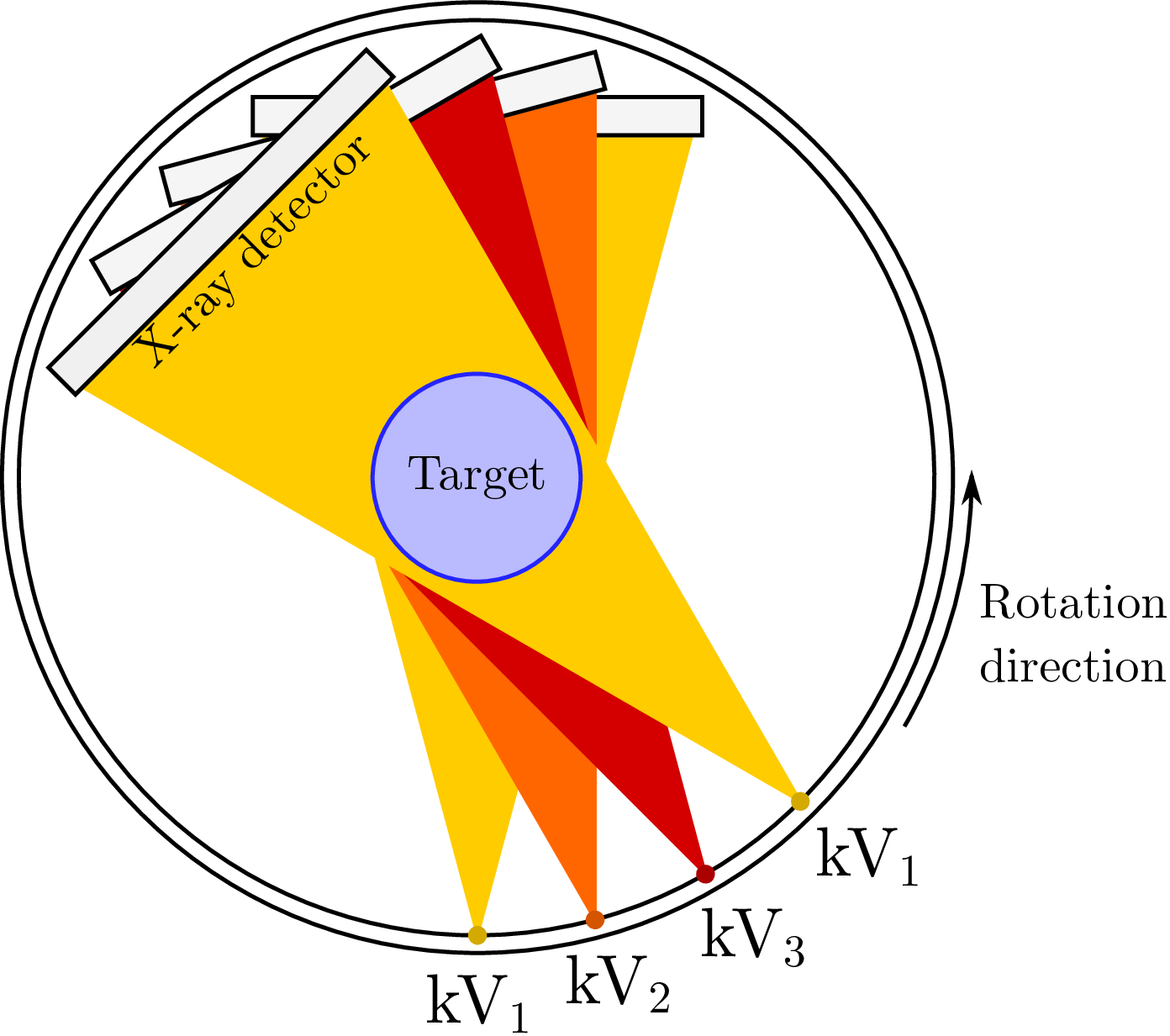}}
\caption{Illustration of a spectrally non-overlapping projection sampling scheme for low dose multi-energy CT imaging. The X-ray source energies $E_k, k=1,2,3$ are denoted by tube voltages 
$\mathrm{kV}_1,\mathrm{kV}_2,\mathrm{kV}_3$.
}
\label{fig:geomfig}
\end{figure}

%We combine the joint reconstruction approaches with a novel low-dose data acquisition protocol. We collect data using a sparse angular sampling, using a different X-ray energy in consecutive steps. See Figure \ref{fig:geomfig} for an illustration. For example with three different energies this protocol drops the radiation dose down to one third compared to the option of measuring all energies at all angles. 
\revmod{We combine the joint reconstruction approach with a low-dose data acquisition protocol. We collect data using a combination of a sparse angular sampling and a method of using a different X-ray energy in consecutive steps similar to \cite{michielsen2018dose}. See Figure \ref{fig:geomfig} for an illustration. For example with three different energies this protocol drops the radiation dose down to one third compared to the option of measuring all the energies at all angles.}

The attenuation images at different energies are structurally similar. Therefore, the combination of non-overlapping sampling and joint regularization model should not compromise image quality. In this paper, the proposed combination of joint reconstruction and non-overlapping sampling is evaluated with simulated multi-energy CT data and experimental data from a biological specimen.  

The rest of this paper is organized as follows. The forward modeling of multi-energy CT is explained in Section \ref{sec:ForwardProblem} and the reconstruction methods used in this paper in Section \ref{sec:InverseProblem}. The generation of simulated data and measuring of the experimental data are explained in Section \ref{sec:MaterialsAndMethods}. The results with simulated and experimental measurement data are shown and discussed in Section \ref{sec:Results}. Section \ref{sec:Conclusions} gives the conclusions.

% ============================================
\section{Forward Problem for Multi-Energy CT}
\label{sec:ForwardProblem}

In multi-energy CT, X-ray images of an object are acquired at two or more X-ray energies to obtain information on the energy-dependent attenuation properties of the object for differentiation and identification of materials. Multiple acquisition schemes exist for obtaining spectral CT data, detailed in, e.g., \cite{mccollough2015dual}. Regardless of how the multi-energy data is acquired, the forward model for multi-energy CT is typically based on the linear projection model of conventional CT. 

To define our forward model, we consider a multi-energy CT experiment with $n$ X-ray energies $\{E_k,\ k = 1,\ldots,n\}$. Let $\Omega \subset \mathbb{R}^m$, $m \in \{2, 3\}$ denote the image domain (2D area or 3D volume) and $\mu  (r, E_k): \Omega \mapsto \mathbb{R}_+$ denote the attenuation function for energy $E_k$ with X-ray source intensity $I_{0} (E_k)$, where $r \in \Omega$ are the spatial coordinates. With these notations, the X-ray intensity at a detector element placed behind the object is modelled by the Beer-Lambert law:
\begin{equation}
\label{eq:beer-lambert}
I = I_{0} (E_k) e^{-\int_\text{ray}\mu (r, E_k)ds},
\end{equation}
where the integration path is along the straight ray from the X-ray source to the detector element \cite{buzug2008ct}. By normalizing with the source intensity and taking the logarithm, we obtain the conventional linear
attenuation model
\begin{equation}
\label{eq:logtransform}
\int_\text{ray}\mu(r, E_k)ds = -\ln\frac{I}{I_{0}(E_k)}.
\end{equation}

\begin{figure}
  \centerline{
  \includegraphics[height=5cm]{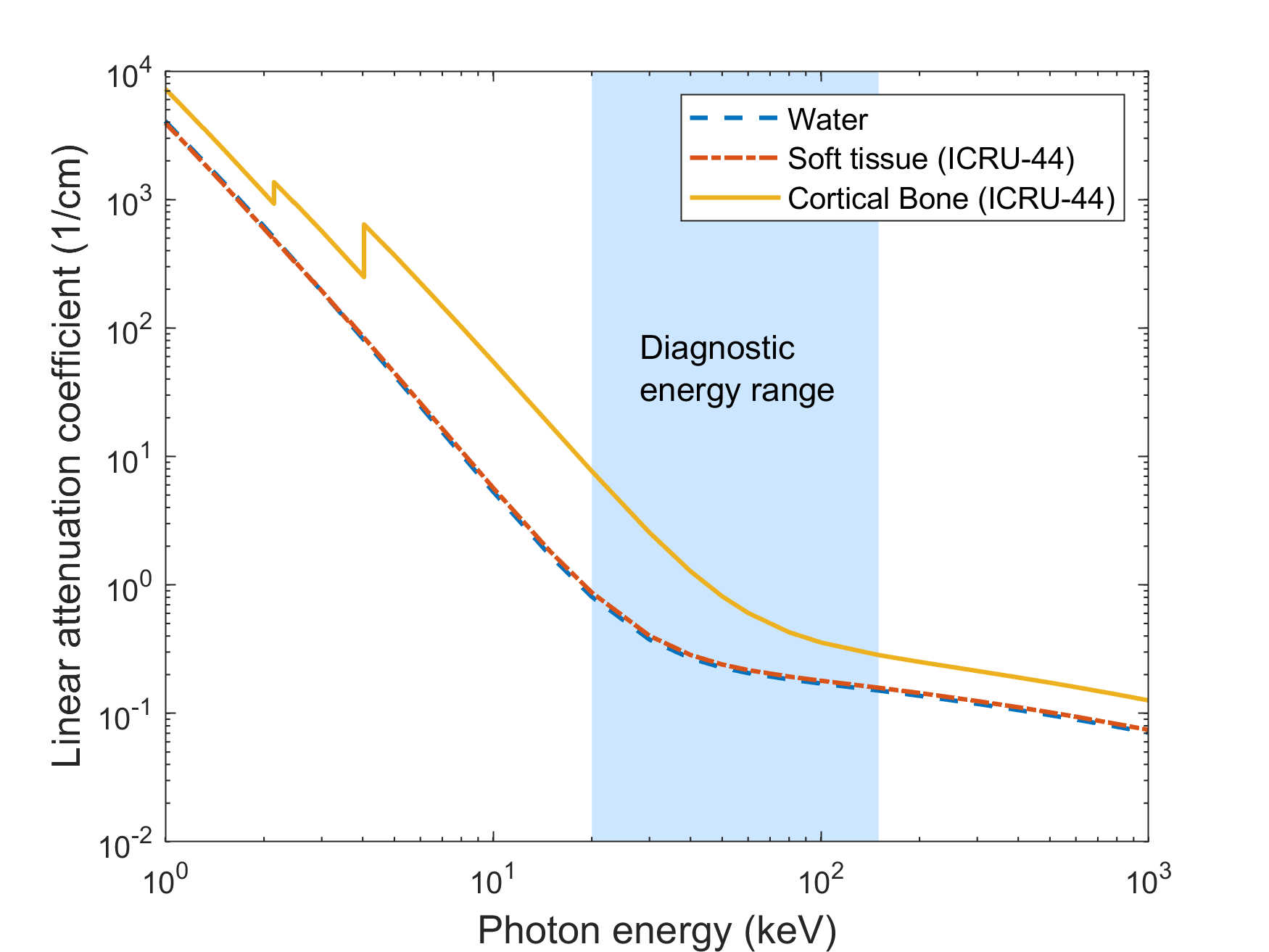}
  \includegraphics[height=5cm]{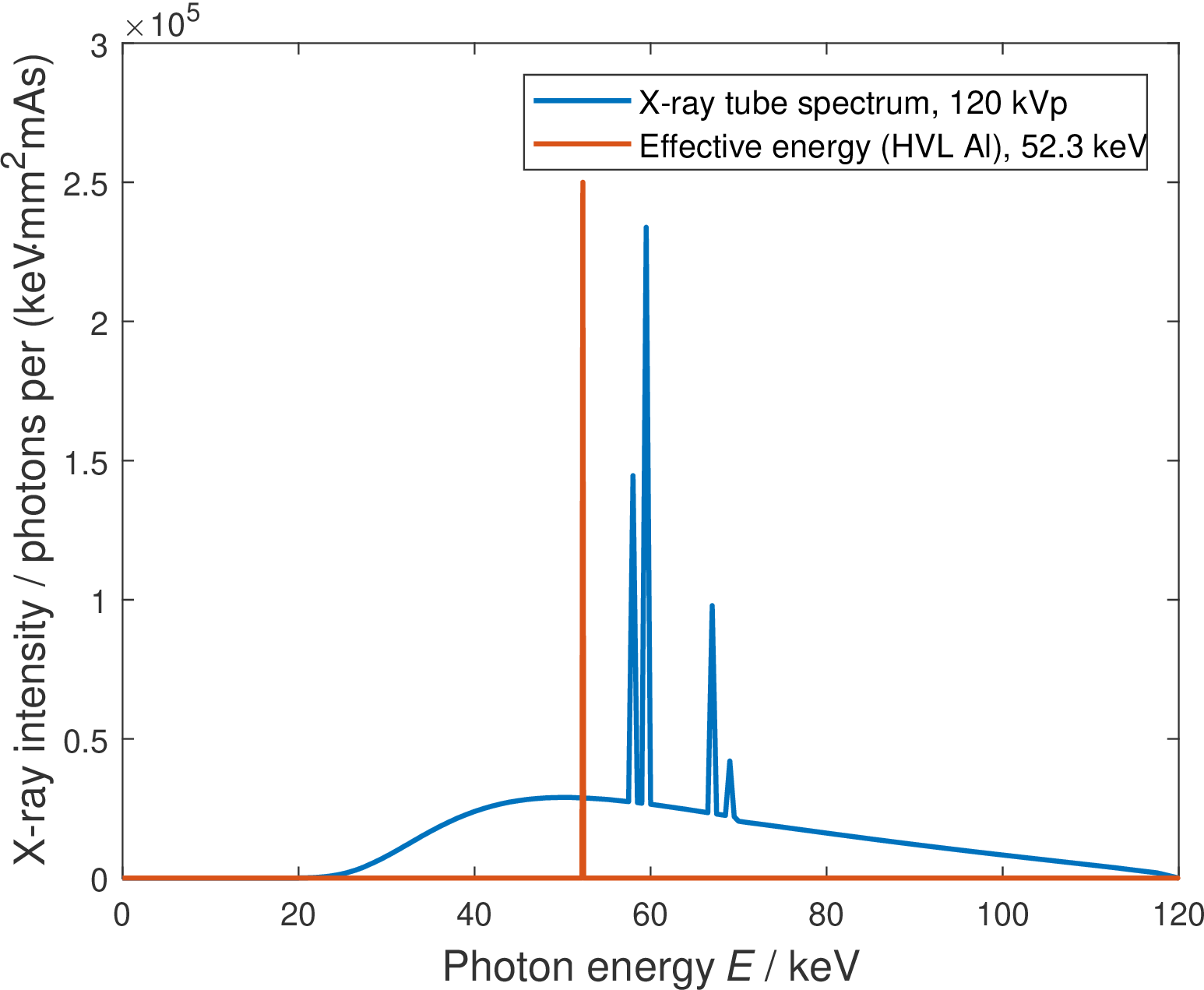}
 }
  \caption{Left: The X-ray attenuation coefficients of water, soft tissue, and cortical bone as functions of X-ray energy. The raw data was obtained from NIST \cite{nist2004xray}. 
Right: Example of a realistic tungsten X-ray source spectrum with 120 kVp voltage and filtering (2.5 mm Al + 0.2 mm Cu), and monochromatic radiation with the same effective energy. The spectra were computed using the SpekCalc software \cite{poludniowski2009spekcalc}.
  }
  \label{fig:xray_attenuation_graphs}
\end{figure}

We remark that the linear attenuation model \eqref{eq:logtransform} is based on several physical approximations, namely neglecting any scattering effects and the energy dependencies of the attenuation coefficients across the X-ray spectrum for polychromatic sources. \revmod{While approaches such as \cite{osullivan2007alternating, chung2010numerical, elbakri2002segmentation} have been proposed for the polychromatic model, these require solving computationally intensive nonlinear optimization problems.}
%Taking all these factors into account would lead to integro-differential light transport models, which would be computationally infeasible.
%
Figure \ref{fig:xray_attenuation_graphs} shows on the left examples of the energy-dependencies of the X-ray coefficients of three materials of medical relevance, and on the right a realistic example of an energy spectrum of an X-ray source.
In principle, when only the scattering effects are neglected, an energy dependent ray propagation based attenuation model would be
\begin{equation}
\label{eq:beer-lambert_polychromatic}
I = \int\limits_0^{E_\text{max}} I_0 (E) e^{-\int_\text{ray}\mu(r,E)ds}dE.
\end{equation}
where $E_\text{max}$ is the maximum energy in the X-ray source spectrum, leading to a non-linear problem for the X-ray tomography. While the non-linear model could in principle be used, it would require accurate knowledge of the input spectrum $I_0(E)$ and either a large number of different measured energies (input spectra) or a spectrally resolved measurement in an attempt to recover $\mu (r,E)$ with a high energy resolution.   
In practice, however, the projection measurements are typically taken only with a few energies (X-ray spectra) and the measurement for each energy is approximated by the linear model \eqref{eq:logtransform} where $E_k$ is thought of as an effective energy for the respective X-ray source spectrum. The effective energy of a polychromatic source is defined as the energy of a monochromatic beam that has the same half-value layer (HVL) as the polychromatic x-ray beam in a given material, usually aluminum or copper \cite{mccketty1998energy}. The use of the linear approximation, however, often results in problems in image reconstruction, such as beam-hardening and metal artifacts \cite{buzug2008ct}. 

For a computational implementation, the forward model \eqref{eq:logtransform} that we use must be appropriately discretized. By concatenating all the X-ray projection data that is measured for the  
(effective) energy $E_k$ into a vector $y_k \in \mathbb{R}^{M_k}$, where $M_k$ is the number of data points for energy $k$, the forward model for the projection data becomes
\begin{equation}
  \label{equ:MeasModel}
  y_k = A_k x_k + e_k
\end{equation}
where $x_k \in \mathbb{R}^N$ is the vectorized representation of the (spatial) discretization of the unknown attenuation image $\mu(r,E_k)$ with energy $E_k$ and $N$ is the number of pixels/voxels, $e_k$ models the measurement noise, and $A_k$ is the matrix which implements the discretization of the projection model
\eqref{eq:logtransform}
for the particular projection geometry and projection angles used with
energy $E_k$. The whole multi-energy CT experiment consist of a set
$\{ y_k, k = 1,\ldots,n\}$ data vectors of the form \eqref{equ:MeasModel}.

We remark that the number $M_k$ of data points for different energies in the joint reconstruction can be freely chosen and could be different for each measured energy (e.g. using fewer measurements for the higher energies).

% ===================================================================
\section{Inverse Problem in Multi-Energy CT}
\label{sec:InverseProblem}

In this section, we discuss the joint reconstruction approach and the joint regularization models that are employed in this study. However, first we will briefly review unregularized reconstruction and total variation (TV) regularized reconstruction for (single energy) CT since they will be employed as references for the joint reconstructions in the numerical examples.

% -------------------------------------------------------------------
\subsection{Least Squares (LS) Reconstruction}

When densely collected projection data for each energy is available, the image reconstructions are typically carried out by using classical methods such as the filtered backprojection (FBP) \cite{natterer1986mathematics}, Kaczmarz iterations such as the algebraic reconstruction technique (ART) \cite{natterer1986mathematics}, or iterative techniques which minimize the least squares (LS) data misfit
\revmod{
\begin{equation}
  \label{equ:FNoPrior}
  x_k = \arg \min_{x_k \geq 0}\left\{|| y_k - A_k x_k ||^2\right\} \regTag{No prior}
\end{equation}
}
%\begin{equation}
%  \label{equ:FNoPrior}
%  F(x_k) = || y_k - A_k x_k ||^2 \regTag{No prior}
%\end{equation}
where $y_k$ is the projection data (using single energy $k$), $A_k$ is the system matrix (or projection matrix) and $x_k$ is the unknown (vectorized) image. 

In this paper, the LS estimate is used as one reference estimate for the joint reconstruction, giving a reference of an estimate which is obtained from the given data $y_k$ without prior information. 

% -------------------------------------------------------------------
\subsection{Sparsity Promoting Reconstruction in CT}

The ALARA principle for X-ray dose in CT imaging has led to significant efforts in development of reconstruction methods that can provide high quality reconstructions from sparsely collected projection data. In the spirit of the theory of compressed sensing \cite{donoho2006compressed}, the methods are usually based on some type of regularization which promotes sparsity of the unknown image by an $l_1$ norm regularization functional on the image on some domain. The $l_1$ regularization can be based, for example, on wavelets \cite{hamalainen2013sparse}, shearlets \cite{bubba2017shearlet}, or the total variation (TV) which promotes piecewise regular solutions by posing the $l_1$ regularization for the gradient of the unknown image \cite{siltanen2003statistical, kolehmainen2003statistical, sidky2008image, hamalainen2014total}.

Using the framework of regularized least squares minimization, the TV regularized reconstruction can be
%stated as minimization of the objective functional
\revmod{obtained as
\begin{equation}
  \label{equ:FSingle}
  x_k = \arg \min_{x_k \geq 0} \left\{|| y_k - A_k x_k ||^2 + \gamma_k TV(x_k)\right\} \regTag{TV}
\end{equation}
}
%\begin{equation}
%  \label{equ:FSingle}
%  F(x_k) = || y_k - A_k x_k ||^2 + \gamma_k TV(x_k) \regTag{TV}
%\end{equation}
where
\begin{align}
  \label{equ:TV}
  TV(x_k) = \int_{\Omega} \left( ||\nabla x_k||^2 + \beta^2 \right)^{1/2} {\rm d}r
\end{align}
is the (smooth) total variation functional \cite{rudin1992nonlinear}, $\beta > 0$ is a smoothing parameter, $\gamma_k > 0$ is a regularization parameter, and $r$ is the spatial coordinate vector. 

In multi-energy CT, one can reconstruct the images at different energies using the TV regularization separately for each energy. In this paper, this approach is used as the other reference for the joint reconstructions, giving a reference of sparsity promoting reconstructions which are computed independently for each energy without any prior model for the correlation of the images between the energies.

% -------------------------------------------------------------------
\subsection{Joint Reconstruction in Multi-Energy CT}
\label{sec:JointReco}

In joint reconstruction, the central idea is to combine all the data sets $\{ y_1, \ldots, y_n \}$, where $n$ is the number of energies, together with a joint regularization functional into a single problem of reconstructing all the unknown images $\{x_1,\ x_2,\ \ldots,\ x_n \}$ at once. In the regularized least squares framework, %the objective functional for the joint reconstruction can be stated as
\revmod{the reconstructions can be obtained as
\begin{equation}
  \label{equ:FMulti}
  (x_1, \ldots, x_n) = \arg \min_{x_1 \geq 0, \ldots, x_n \geq 0} \left\{ \sum_{k=1}^{n} || y_k - A_kx_k ||^2  + R(x_1, x_2, \ldots, x_n) \right\} 
\end{equation}
}
% (x_1,\ldots,x_n) = \arg \min_{x_1 \geq0, \ldots, x_n \geq 0} \{ \| ...\|\}
%\begin{equation}
%  \label{equ:FMulti}
%  F(x_1,\ldots,x_n) = \sum_{k=1}^{n} || y_k - A_kx_k ||^2  + R(x_1, x_2, \ldots, x_n)
%\end{equation}
where $y_k$ is the measurement vector and $A_k$ is the system matrix corresponding to energy $k$, and
\begin{equation}
  \label{equ:regFunc}
  R(x_1, x_2, \ldots, x_n) = \sum_{k=1}^{n} \gamma_k U(x_k) + \alpha W(x_1, x_2, \ldots, x_n)
\end{equation}
is the regularization functional where $U(x_k)$ is a (spatial) regularization functional acting separately on each of the images $x_k$, $W(x_1, x_2, \ldots, x_n)$ is a joint regularization functional which is used to incorporate prior information across the different energies, and $\gamma_k$ and $\alpha$ are regularization parameters.

The regularization functional $W(x_1, x_2, \ldots, x_n)$ should be formulated to promote feasible \textit{a priori} information between the unknown images at different energies. In multi-energy CT, it is reasonable to assume that the attenuation images at different energies are structurally similar in the sense that if an edge is present with one energy, it is likely to be at the same location and alignment with other energies as well. 

The following subsections introduce the joint regularization functionals used in this paper.

% -------------------------------------------------------------------
\subsubsection{Joint Total Variation}

The joint total variation (JTV) functional was originally developed for denoising and deblurring of the red, green, and blue channels in RGB images \cite{blomgren1998color}. It can be defined as
\begin{align}
  \label{equ:JTV}
  JTV(x_1, x_2, \ldots, x_n) = \int_{\Omega} \left( ||\nabla x_1||^2 + ||\nabla x_2||^2 + \ldots + ||\nabla x_n||^2 + \beta^2 \right)^{1/2} {\rm d}r
\end{align}
and it promotes sparsity of the joint gradient, leading to edge preserving regularization within each image and favoring the locations of edges to be the same in the images $x_k$.

In the JTV regularization, in \eqref{equ:regFunc} we set $U(x_i) = 0$ and\\ $W(x_1, x_2, \ldots, x_n) = JTV(x_1, x_2, \ldots, x_n)$
leading to
\begin{equation}
  \label{equ:JTVreg}
  R(x_1, x_2, \ldots, x_n) = \alpha JTV(x_1, x_2, \ldots, x_n) \regTag{JTV}
\end{equation}
in \eqref{equ:FMulti}.

% -------------------------------------------------------------------
\subsubsection{Linear Parallel Level Sets}

The linear parallel level sets (LPLS) prior was originally used for denoising and demosaicking RGB images \cite{ehrhardt2014vector}. In tomographic context, it has been used in joint reconstruction of PET and MRI images \cite{ehrhardt2014joint}, reconstruction of PET and EIT images with MRI images as side information \cite{ehrhardt2016pet, kolehmainen2019incorporating} and in multi-energy computed tomography \cite{kazantsev2018joint}.

The LPLS prior is based on the idea that level sets can be used to identify the structure in images and that the gradients of an image are always perpendicular to the level sets. Therefore, structural similarity can be measured by measuring the parallelism of the gradients’ orientations in two images. In the context of a minimization problem, this can be formulated as \cite{ehrhardt2016pet}
\begin{align}
  \label{equ:LPLS}
  \nonumber LPLS(x_k, x_j) = \int_{\Omega} \left(\left( ||\nabla x_k||^2 + \beta^2 \right)^{1/2}\left( ||\nabla x_j||^2 + \beta^2 \right)^{1/2} \right.\\
  \left. - \left( |\langle\,\nabla x_k, \nabla x_j\rangle|^2 + \beta^4 \right)^{1/2} \right) {\rm d}r
\end{align}
and in this paper, the computation is done in pairs, so that
\begin{align}
  \nonumber
  LPLS(x_1, x_2, \ldots, x_n)
  = LPLS(x_1,x_2) + LPLS(x_2,x_3) + \\
  \ldots + LPLS(x_n,x_1).
\end{align}

In the LPLS regularization, we set $U(x_i) = 0$ and\\
$W(x_1, x_2, \ldots, x_n) = LPLS(x_1, x_2, \ldots, x_n)$
in \eqref{equ:regFunc}, leading to the regularization functional
\begin{equation}
  \label{equ:LPLSreg}
  R(x_1, x_2, \ldots, x_n) = \alpha LPLS(x_1, x_2, \ldots, x_n) \regTag{LPLS}
\end{equation}

% -------------------------------------------------------------------
\subsubsection{Spectral Smoothness Regularization}

A straightforward way to promote structural similarity is to use smoothness regularization in the spectral (energy) dimension. Spectral smoothness regularization can be obtained simply by penalizing differences between the images $\{ x_k \}$. In this paper, we consider a spectral smoothness regularization term based on the first finite difference but also higher order finite differences could be used. Somewhat similar smoothness penalties have been used for example in dynamic magnetic resonance imaging \cite{rasch2018dynamic} and X-ray fluorescence tomography \cite{gursoy2015hyperspectral}.

The first finite difference can be used for regularization with
\begin{align}
  \label{equ:D1}
    D1(x_k,x_{k+1}) = w_k \int_{\Omega} ||x_{k+1} - x_{k}||^2 {\rm d}r
\end{align}
and it favors images that are close to being identical. The parameter $w_k$ in \eqref{equ:D1} basically corresponds to the reciprocal of the (spectral) distance (energy difference) between the images $x_{k+1}$ and $x_k$. However, in cases where one might have prior information about the spatial variations in the expected magnitudes of the changes between the images $x_{k+1}$ and $x_k$, the parameter $w_k$ could be made spatially distributed for construction of a spatially weighted regularization model for the spectral smoothness. 
For more than two images $x_i$ we can write
\begin{align}
  \label{equ:D1More}
    D1(x_1,x_2,\ldots,x_n) = D1(x_1,x_2) + D1(x_2,x_3) + \ldots + D1(x_{n-1},x_{n}).
\end{align}
In this paper, we set the weights $w_k \equiv 1$, and use the regularization functional
\begin{align} \label{D1reg}
  R(x_1,x_2,\ldots,x_n) = \alpha D1(x_1,x_2,\ldots,x_n)  \regTag{D1}
\end{align}
for D1 regularization.

Notice that the D1 regularizer imposes a penalty only for the differences of the pixel values in the energy direction. When using highly sparse projection data, this lack of spatial constraint can result in spatially noisy images. In such cases, an additional spatial regularization term, such as total variation (\ref{equ:TV}), can be used to promote regularity within each of the images. This can be done by choosing $U(x_i) = TV(x_i)$ in (\ref{equ:regFunc}), leading to
\begin{align}
  R(x_1,x_2,\ldots,x_n) =  \sum_{k=1}^{n} \gamma_k TV(x_k) + \alpha D1(x_1, x_2, \ldots, x_n)  \regTag{D1+TV} 
\end{align}
for the D1+TV regularization.

% -------------------------------------------------------------------
\subsubsection{Structural Similarity Index Based Regularization}

A new joint regularization model which favors structural similarity can be formulated by considering the structural similarity (SSIM) index \cite{wang2004image,wang2009mean} which is an image metric that assesses the similarity of two images by using three characteristics of the input images: luminance, contrast, and structure. Here we consider only the structure part which associates the unit vectors $(x_1-\mu_{x_1})/\sigma_{x_1}$ and $(x_2-\mu_{x_2})/\sigma_{x_2}$ with the structure of the two images and uses their inner product to measure the structural similarity, resulting in
\begin{align}
  \label{equ:Si}
  \hat{S}(x_1,x_2) = \frac{\sigma_{x_1x_2} + C}{\sigma_{x_1} \sigma_{x_2} + C}
\end{align}
where $\sigma_{x_1x_2}$ is the cross correlation, $\sigma_{x_1}$ and $\sigma_{x_2}$ are the standard deviations, and $\mu_{x_1}$ and $\mu_{x_2}$ are the expected values of $x_1$ and $x_2$, and $C$ is a small constant introduced to avoid division by zero. Geometrically, the cross correlation $\sigma_{x_1x_2}$ corresponds to the cosine of the angle between the vectors $x_1 - \mu_{x_1}$ and $x_2 - \mu_{x_2}$, and thus (\ref{equ:Si}) is maximized when these vectors point in the same direction and minimized when they point in opposite directions.

As with the SSIM index, it is possible to compute an image showing the similarity in structure of two images. Following the procedure for computing a SSIM map \cite{wang2004image}, but only for the structure part (\ref{equ:Si}), the cross correlations and standard deviations, as well as $\hat{S}$ itself, are computed locally using a 11x11 window that moves pixel by pixel over the entire image. If we let $\mathcal{N}_i$ denote the set of indices for the 11x11 window for pixel $i$ in the image lattice, and let $w = \{w_k | k \in \mathcal{N}_i \}$ denote the local weights obtained from a circular symmetric Gaussian weighting function with standard deviation of 1.5, local weighted statistics for window $i$ can be computed as
\begin{align}
  \label{equ:sx}
  (\sigma_{x_k})_i =& \left( \sum_{j \in \mathcal{N}_i}w_j((x_k)_j - \mu_{x_k})^2 \right)^{\frac{1}{2}}\\
  \label{equ:mx}
  (\mu_{x_k})_i =& \sum_{j \in \mathcal{N}_i}w_j(x_k)_j\\
  \label{equ:sxx}
  (\sigma_{x_kx_p})_i =& \sum_{j \in \mathcal{N}_i}w_j\big((x_k)_j - \mu_{x_k}\big)\big((x_p)_j - \mu_{x_p}\big)
\end{align}
where $(x_k)_j$ is the $j$th element of $x_k$. The obtained pixel-by-pixel local $\hat{S}$ form the structural similarity map with larger values indicating similarity and smaller values indicating structural differences. Now we can use the mean of the local $\hat{S}$ values
\begin{align}
  \label{equ:barS}
  \bar{S}(x_k,x_p) = \frac{1}{N}\sum_{i=1}^{N} (\hat{S}(x_{k},x_{p}))_i
\end{align}
where $N$ is the number of pixels in the images and
$(\hat{S}(x_{k,i},x_{p,i}))_i$ is the value of $\hat{S}$ in pixel $i$, as a scalar measure of the structural similarity which is maximized with the value 1 when the images $x_k$ and $x_p$ are identical. 
Now the regularization term in (\ref{equ:FMulti}) can be chosen to be some decreasing function of all the $\bar{S}(x_k,x_{k+1})$ which has minimum when the image pairs are identical. In this paper, we use a simple choice
\begin{align}
  \label{equ:S}
  S(x_1, x_2, \ldots, x_n) = \frac{1}{\bar{S}(x_1, x_2, \ldots, x_n)}
\end{align}
where
\begin{align}
  \bar{S}(x_1, x_2, \ldots, x_n) = \bar{S}(x_1,x_2) + \bar{S}(x_2,x_3) + \ldots + \bar{S}(x_n,x_1)
\end{align}
computes the structure parts in pairs.

When employing the structural similarity regularization alone, we set $U(x_i) = 0$ and use
\begin{align} \label{Sreg}
  R(x_1,x_2,\ldots,x_n) = \alpha S(x_1, x_2, \ldots, x_n)  \regTag{S}
\end{align}
as the regularization functional in (\ref{equ:FMulti}) for S regularization.

Similarly as with the spectral smoothness regularization, the lack of penalty for spatial regularity within each channel in the structural regularization term (\ref{equ:S}) can result in spatially noisy reconstructions when using sparse projection data. In such cases, an additional spatial regularization term, such as total variation (\ref{equ:TV}), can be used to promote spatial regularity of the reconstructions. In this paper, we use $U(x_i) = TV(x_i)$ in (\ref{equ:regFunc}), leading to
\begin{align}
   R(x_1,x_2,\ldots,x_n) =  \sum_{k=1}^{n} \gamma_k TV(x_k) + \alpha S(x_1, x_2, \ldots, x_n) \regTag{S+TV}
\end{align}
for the S+TV regularization.

% ===================================================================
\section{Materials and Methods}
\label{sec:MaterialsAndMethods}

In this section, we present the simulated and measured X-ray projection data that were used to test the different regularization models utilized in multi-energy CT reconstruction. The datasets were generated using dense angle sampling, and could then later be appropriately subsampled to simulate sparse-angle acquisition schemes.

% -------------------------------------------------------------------
\subsection{Simulated Data}
\label{sec:SimuData}

Simulated X-ray projection data was generated using idealized, monochromatic sources in a parallel beam geometry. For this, a computational anthropomorphic 2D phantom was created using the XCAT software \cite{segars2008xcat,segars2010xcat}. The phantom consisted of a single cross-sectional slice of an adult human male chest. A calcified coronary artery was placed in the heart of the phantom. We simulated three sets of noise-free projections using 40 keV, 80 keV, and 120 keV X-rays. Each set consisted of 360 projections taken evenly across \ang{180} at \ang{0.5} intervals. The X-ray detector was a line detector consisting of 729 elements, with a pixel size of 0.875 mm. To avoid committing inverse crime, the projections were generated using XCAT's own CT Projector tool.

The noisy realizations of the simulated measurements for each energy were obtained by adding Gaussian zero mean random noise with standard deviation equal to one percent of the maximum amplitude of the measurements with that energy to the simulated noise-free measurements. 

The reconstruction grid size was $512 \times 512$ pixels. The forward operators for the model $Ax_k = y_k$ were created using the ASTRA Toolbox \cite{van2016fast, van2015astra, astrawebpages}. 

% -------------------------------------------------------------------
\subsection{Experimental Data}
\label{sec:ExpData}

Experimental data was measured at the Department of Physics, University of Helsinki, using a cone-beam micro-CT scanner with an end-window tube and a tungsten target (GE Phoenix nanotom 180 NF). The chest of a small bird (the common quail, \emph{Coturnix coturnix}) was used as a test phantom, as it contains multiple different tissue types as well as fine details arising from the bone structure. The bird phantom was imaged using three different X-ray tube settings in the same geometry. Multiple frames were averaged for each projection in order to increase signal-to-noise ratio. The imaging geometry is specified in Table \ref{table:geometry} and the settings for each effective energy are specified in Table \ref{table:energy}. 2D sinograms were created using the central plane of the cone-beam projections, in which the geometry reduces to a fan-beam geometry.

The reconstruction grid size was $512 \times 512$ pixels with a pixel size of 120 \si{\micro m}, and the forward operators for the imaging model were created using the ASTRA Toolbox.

\begin{table}
\caption{Imaging geometry used for collecting the experimental data.
\label{table:geometry}
}
    \begin{center}
    \begin{tabular}{ l  l }
        \hline
        Parameter & Value \\
        \hline
        \hline
      Focus-center distance & 252 mm \\
      Focus-detector distance & 420 mm \\
        Magnification & 5/2 \\
        Detector pixel size & 0.200 mm \\
        Effective pixel size & 0.120 mm \\
        Projection size & 552 $\times$ 576 pixels \\
    Angular range & \ang{360}\\
        \#projections & 720 \\
        \hline
      \end{tabular}
  \end{center}
\end{table}

\begin{table}
  \caption{Energy-specific settings used for collecting the experimental data.}
    \label{table:energy}
  \begin{center}
      \begin{tabular}{ l  l  l  l  l  l}
      \hline
    Energy & $U$ (kV) & Filtration & $I$ (\si{\micro A}) & Exposure time (ms) & Frame averaging\\
        \hline
      \hline
      $E_1$ & 50 & None & 300 & 125 & 4\\
      $E_2$ & 80 & 1 mm Al & 180 & 125 & 4\\
      $E_3$ & 120 & 0.5 mm Cu & 120 & 250 & 4\\
      \hline
      \label{tabular:measurementSpecs}
      \end{tabular}
  \end{center}
\end{table}

% -------------------------------------------------------------------
\subsection{Image Fidelity Measures}
\label{sec:Fidelity}
To quantify the quality of reconstructions in the simulation test case, we use the root mean square error (RMSE) and the mean structural similarity (MSSIM) index \cite{wang2004image}.

The root mean square error can be written as
\begin{equation}
  \label{equ:RMSE}
  \textrm{RMSE}(x_k) = \sqrt{\frac{1}{N} \sum_{i=1}^{N} ((x_k)_i - (x_{\textrm{REF,k}})_i)^2}
\end{equation}
where $(x_k)_i$ denotes the $i$th pixel of the solution $x$ corresponding to energy $k$ and $x_{\textrm{REF},k}$ is the reference image corresponding to energy $k$.

The structural similarity (SSIM) index \cite{wang2004image} aims to measure the perceived similarity in two images, $x_1$ and $x_2$. It assesses the luminance $l(x_1,x_2)$, contrast $c(x_1,x_2)$ and structure $s(x_1,x_2)$ and can be written as
\begin{align}
  \label{equ:SSIM}
  \nonumber \textrm{SSIM} &= l(x_1,x_2) c(x_1,x_2) s(x_1,x_2) \\
  &= \left( \frac{2\mu_{x_1} \mu_{x_2} + C_1}{\mu_{x_1}^2 + \mu_{x_2}^2 + C_1} \right)
    \left( \frac{2\sigma_{x_1}\sigma_{x_2} + C_2}{\sigma_{x_1}^2 + \sigma_{x_2}^2 + C_2} \right)
    \left( \frac{\sigma_{x_1x_2} + C_3}{\sigma_{x_1} \sigma_{x_2} + C_3} \right)
\end{align}
where $\mu$ is the mean, $\sigma$ is the standard deviation, $\sigma_{x_1x_2}$ is the cross-correlation and $C_1$, $C_2$ and $C_3$ are small constants used to avoid division by zero. The mean SSIM index is obtained by first computing the SSIM index locally for each pixel using a 11x11 pixel window whilst also using a Gaussian weighting function for the statistics, as in (\ref{equ:sx} -- \ref{equ:sxx}), and then the final index is obtained as the mean of these local SSIM indices. For a reconstructed image $x_k$ and a reference image $x_{\textrm{REF},k}$, this can be written as
\begin{align}
  \label{equ:MSSIM}
  \textrm{MSSIM}(x_k,x_{\textrm{REF},k}) = \frac{1}{N}\sum_{i=1}^{N} (\textrm{SSIM}(x_{k}, x_{\textrm{REF},k}))_i
\end{align}
where $(\textrm{SSIM}(x_{k},x_{\textrm{REF},k}))_i$ is the local value of SSIM for pixel $i$ in the image lattice.

As the anthropomorphic XCAT target phantom that was used as the 
basis for the projection data does not translate well into attenuation images that could be used as ground truth reference images, the ground truth reference images $x_{\textrm{REF},k}$ for the simulation test case were obtained by computing full {\em noiseless data} (360 angles for each energy) filtered backprojection (FPB) reconstructions using the ASTRA Toolbox \cite{van2016fast,van2015astra,astrawebpages} FBP algorithm.

For the test case with experimental data, the reference images $x_{\textrm{REF},k}$ were obtained by the 'No Prior' method in \eqref{equ:FNoPrior} applied to 720 projections for each measured energy.

% -------------------------------------------------------------------
\subsection{Selection of the Regularization Parameters}

For the simulation test case, the regularization parameters $\gamma_k$ and $\alpha$ in (\ref{equ:FSingle}) and (\ref{equ:FMulti}) and $\beta$ for each regularization term were selected by computing a series of reconstructions with varying $\gamma_k$, $\alpha$ and $\beta$. Then the mean structural similarity index (\ref{equ:MSSIM}) was computed for each reconstructed image using the full data FPB reconstructions as the reference images $x_{\textrm{REF},i}$, and the $\gamma_k$, $\alpha$ and $\beta$ corresponding to the highest structural similarity were selected for the test cases. A similar procedure was used in the experimental data test case but the identification of best reconstructions was done with visual comparison. The selected $\gamma_k$ and $\alpha$ for each regularization term are given in Table \ref{tab:alphas}. For $\beta$, the value of 1e-6 was used for the simulated data test case and 1e-9 for the experimental data test case. \revmod{These values are small compared to the magnitudes of the gradients in Equations (\ref{equ:TV}), (\ref{equ:JTV}) and (\ref{equ:LPLS}), resulting to an acceptable level of smoothing in the reconstructions.}

\begin{table}[htbp]
  \footnotesize
  \begin{center}
    \caption{Selected $\alpha$ and $\gamma_k$ parameters for the different regularization terms for the simulated data test case ($\alpha_{\textrm{SIM}}, \gamma_{k,\textrm{SIM}}$) and for the experimental data test case ($\alpha_{\textrm{EXP}}, \gamma_{k,\textrm{EXP}}$). The three values of $\gamma_k$ are for the three different tube energies with indices $k=1,2,3$.}
    \label{tab:alphas}
    \begin{tabular}{llllll}
      \hline
      Prior & Acronym & $\alpha_{\textrm{SIM}}$ & $\gamma_{k,\textrm{SIM}}$ & $\alpha_{\textrm{EXP}}$ & $\gamma_{k,\textrm{EXP}}$ \\
      \hline
      No prior & No prior & - & - & - & - \\
      Total variation & TV & - & 2.25, 0.8, 0.7 & - & 1.0, 0.5, 0.25\\
      Joint total variation & JTV & 2.25 & 0 & 1 & 0\\
      Linear parallel level sets & LPLS & 4000 & 0 & 6000 & 0\\
      First difference & D1 & 400 & 0 & 100 & 0\\
      Structural & S & 4e6 & 0 & 1e6 & 0\\
      First difference + TV & D1+TV & 50 & 1.125, 0.4, 0.35 & 70 & 0.5, 0.25, 0.125\\
      Structural + TV & S+TV & 0.75e5 & 1.125, 0.4, 0.35 & 0.5e6 & 0.5, 0.25, 0.125\\
      \hline
    \end{tabular}
  \end{center}
\end{table}

% -------------------------------------------------------------------
\subsection{Solution of the image reconstruction problems}
The minimization problems (\ref{equ:FNoPrior}), (\ref{equ:FSingle}) and (\ref{equ:FMulti}) were solved using a Polak-Ribi\`ere nonlinear conjugate gradient method \cite{polak1969note} where the search direction was reset to the deepest descent direction whenever the Polak-Ribi\`ere $\beta^{PR}$ parameter went negative. The step lengths were chosen using a golden section line search and the non-negativity constraints were implemented using projection. The iteration was terminated when the change in the minimized function or the norm of the change in the parameters was less than 1e-9, or after 512 iterations. The maximum number of iteration limit was reached only when computing the \textit{no prior} solution (\ref{equ:FNoPrior}). \revmod{Using a smaller termination criterion or a larger maximum iteration limit did not lead to noticeable changes in the reconstructions.} \revmod{We remark that the nonlinear Polak-Ribi\`ere conjugate gradient algorithm does not possess global convergence results under the present setup where we minimize different non-linear functionals under non-negativity constraints for the solution. However, the algorithm implementation for the different regularization models was tested  with simulated data with various targets, with and without simulated measurement errors, and with full and sparse measurement data. The minimization algorithm converged to a (possibly local) minimum and produced feasible reconstructions in all the test cases.}

% ===================================================================
\section{Results}
\label{sec:Results}

In this section, reconstructions computed using the joint reconstruction approaches in Section \ref{sec:JointReco} are presented and compared to the full data reference images and to the single-energy reference methods (\ref{equ:FNoPrior}) (no prior) and \eqref{equ:FSingle} (TV) where attenuation at each energy is reconstructed separately. The first subsection introduces the multi-energy projection geometry, the second subsection shows results computed with simulated data and the third subsection shows results computed with experimental measurement data.

% -------------------------------------------------------------------
\subsection{Multi-Energy Projection Geometry}

The principle of the low dose multi-energy projection geometry is illustrated in Figure \ref{fig:geomfig} for an experiment with three different energies, denoted by the X-ray tube voltages $\mathrm{kV}_1$, $\mathrm{kV}_2$ and $\mathrm{kV}_3$ in the picture. Instead of sampling all the projection directions with all the energies, the directions are divided to non-overlapping sets of interleaved projection directions making the projection geometry different for each of the used source energies. This type of imaging setup could be realistically implemented in a clinical CT scanner using a rapid-kVp switching scheme.

In this study, the non-overlapping low dose data sets were obtained by retrospective subsampling of the full angle data sets with three different source energies. In both the simulation and experimental data cases, a subset of 90 equally spaced projection angles were chosen from the whole angular span of the projection geometry. This subset was then further subsampled to three subsets of 30 angles, leading to 30 equally spaced projection directions for each energy, see Figure \ref{fig:geomfig} and Table \ref{tab:angles}. For each energy, the projection data corresponding to the set of 30 angles was used in the computations.

\begin{table}[htbp]
  \footnotesize
  \begin{center}
    \caption{Measurement angles used for energies $E_1$, $E_2$ and $E_3$ in the simulated data case (left) and experimental data case (right).}
    \label{tab:angles}
    \begin{tabular}{lllllll}
          $E_1$ & $E_2$ & $E_3$ & & $E_1$ & $E_2$ & $E_3$ \\
      \hline
         0   &  2  &   4& &  0   &  4  &   8 \\
         6   &  8  &  10& &  12  &  16 &   20\\
        12   & 14  &  16& &  24  &  28 &   32\\
        $\vdots$ & $\vdots$ & $\vdots$ & & $\vdots$ & $\vdots$ & $\vdots$\\
       162 &  164 &  166& &   324 &  328 &  332\\
       168 &  170 &  172& &   336 &  340 &  344\\
       174 &  176 &  178& &   348 &  352 &  356\\
      \hline
    \end{tabular}
  \end{center}
\end{table}

% -------------------------------------------------------------------
\subsection{Simulated Data}

\begin{figure}[p]
  \includegraphics[height=0.875\textheight]{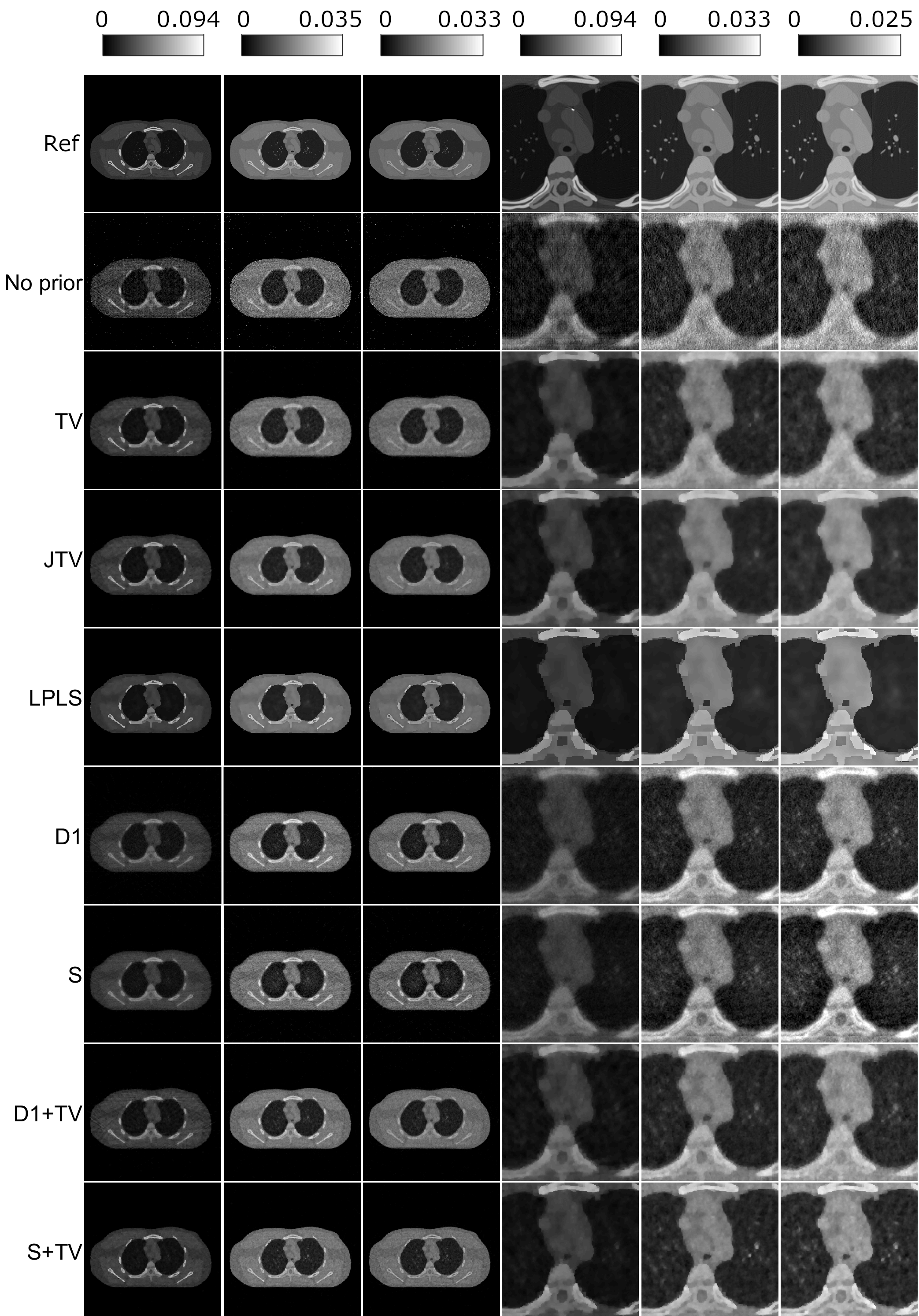}
  \caption{Reconstructed images for the simulation test case using 30 projection directions at each energy.}
  \label{fig:SIM30}
\end{figure}

\begin{figure}
  \includegraphics[width=0.275\paperwidth]{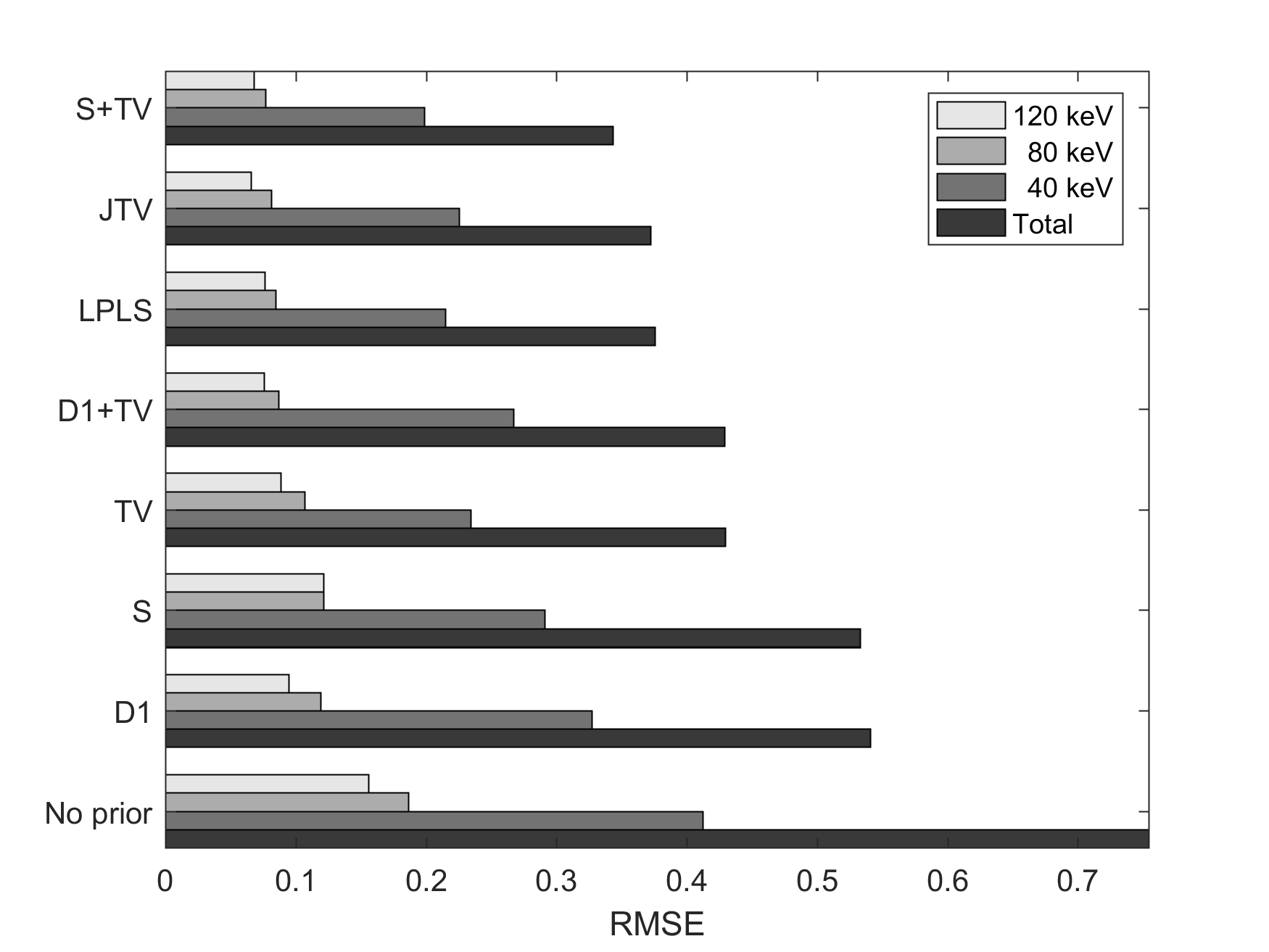}
  \includegraphics[width=0.275\paperwidth]{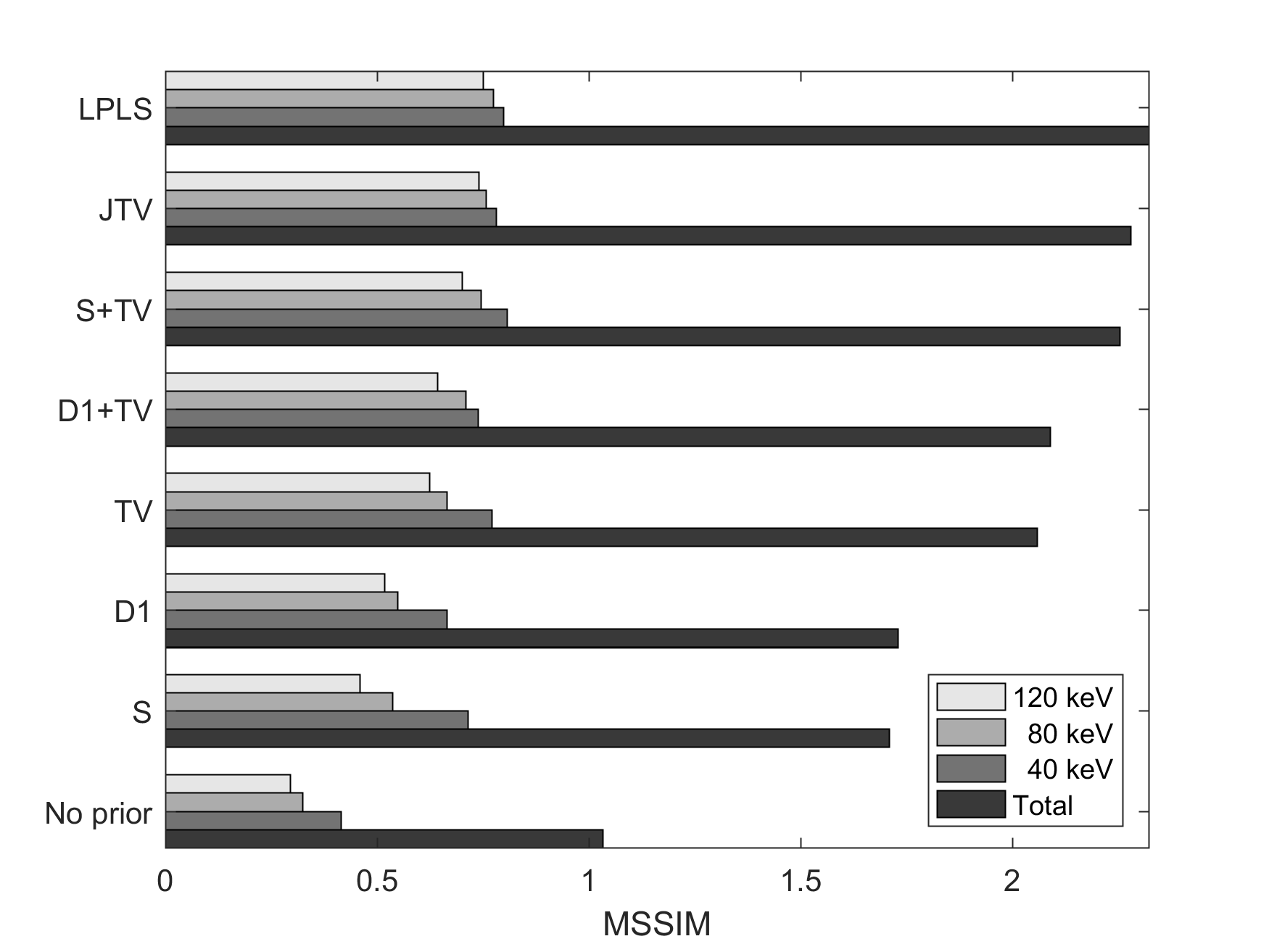}
  \caption{The RMS errors (left) and the mean SSIM indices (right) for the different reconstruction approaches using 30 projection directions at each energy.}
  \label{fig:SIM30Metrics}
\end{figure}

The reconstructed images for the simulated data test case are shown in Figure \ref{fig:SIM30}. The first three columns show the reconstructions for energies 40, 80 and 120 keV, respectively, and the next three columns show a region of interest from the corresponding reconstructions. 
Figure \ref{fig:SIM30Metrics} shows the RMSE and MSSIM fidelity metrics for the reconstructions. 

All the joint reconstructions (JTV, LPLS, D1, S, D1+TV, S+TV) in Figure
\ref{fig:SIM30} are visually better than the independent single energy reconstructions (No prior, TV) in the sense that more of the details that are present in the full data reference images can be detected in the reconstructions. This could be expected since the sub-problems with different energies and projection directions are coupled to each other in the joint reconstruction approaches via the joint regularization functional. 
The (S+TV) reconstructions have the best fidelity in the RMSE measure and the (LPLS) reconstructions in the MSSIM measure, see Fig \ref{fig:SIM30Metrics}. One interesting feature in the reconstructions is the difference in the amount of details in the reconstructions with the (JTV, LPLS) models compared to the reconstructions with the (D1, S) models. The JTV and LPLS models promote spatially piecewise regular solutions by posing different penalties for the sparsity of the image gradients. Consequently, the reconstructions lack some of the fine details but on the other hand have a very low level of spatial noise and very good RMSE and MSSIM metrics. 
In contrast, the reconstructions with the (D1, S) models show more details but are somewhat noisy, which is also reflected in the fact that the (D1, S) models have worse RMSE and MSSIM metrics than the (TV) model. This behavior is due to the fact that the (D1, S) models are based on promoting structural similarity in the energy direction without any penalty on spatial regularity. A good combination between the spatial details and noise level is reached when the (D1, S) models are used in combination with the TV regularization applied to each of the images for promoting spatial regularity (D1+TV, S+TV).

\begin{figure}[p]
  \centering
  \includegraphics[width=0.9\textwidth]{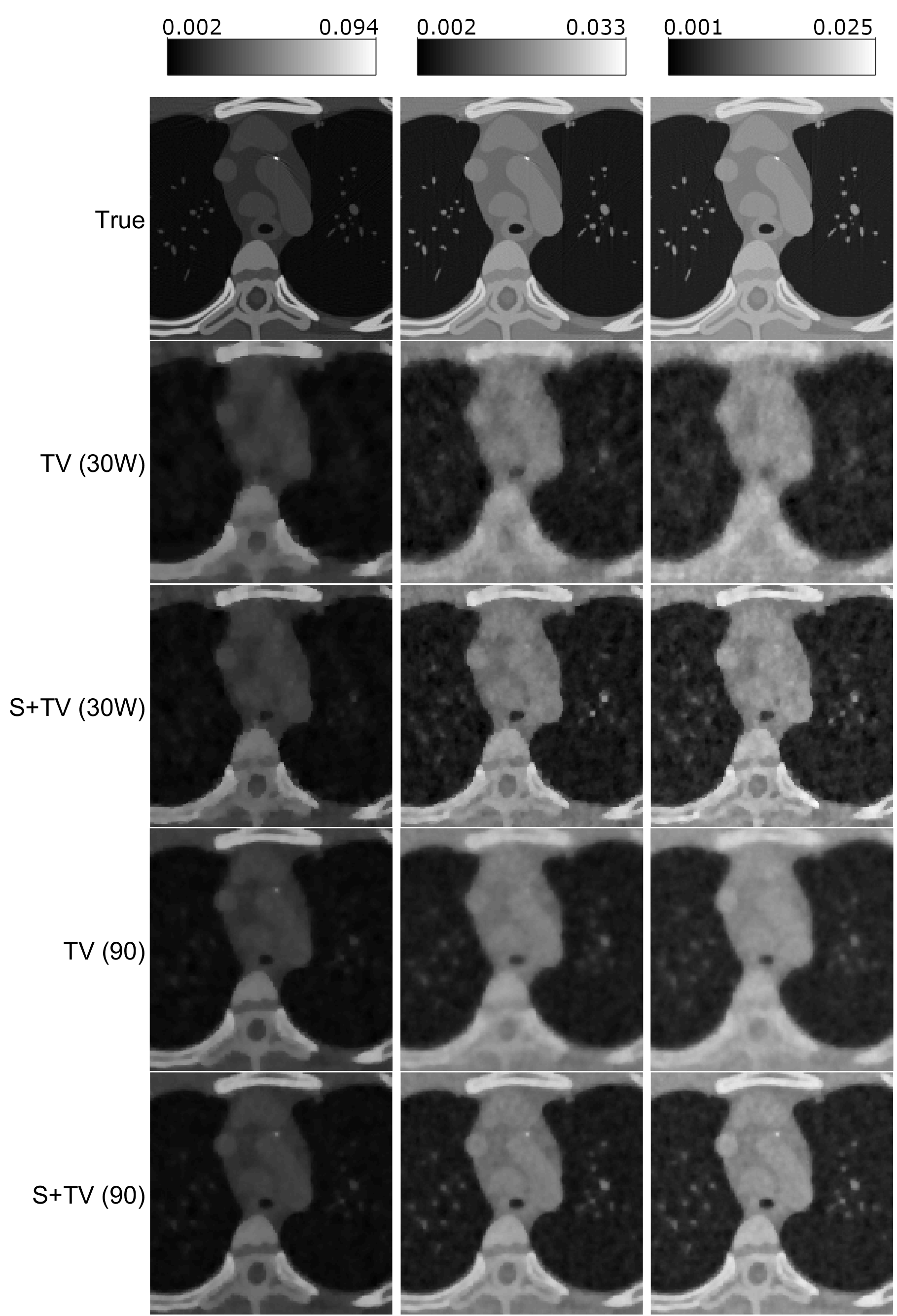}
  \caption{Comparison of TV and S+TV reconstructions. Results labeled METHOD(90) are based on overall of $3\times 90 = 270$ projection images using the same 90 directions for each energy. Results labeled METHOD(30W) are based on overall of 90 projections using the non-overlapping 30+30+30 directions (the same geometry that is used in Figure \ref{fig:SIM30}). Each image shows a region of interest of the full reconstruction.}
  \label{fig:90vs30images}
\end{figure}

\begin{figure}
  \centering
  \includegraphics[width=0.75\textwidth]{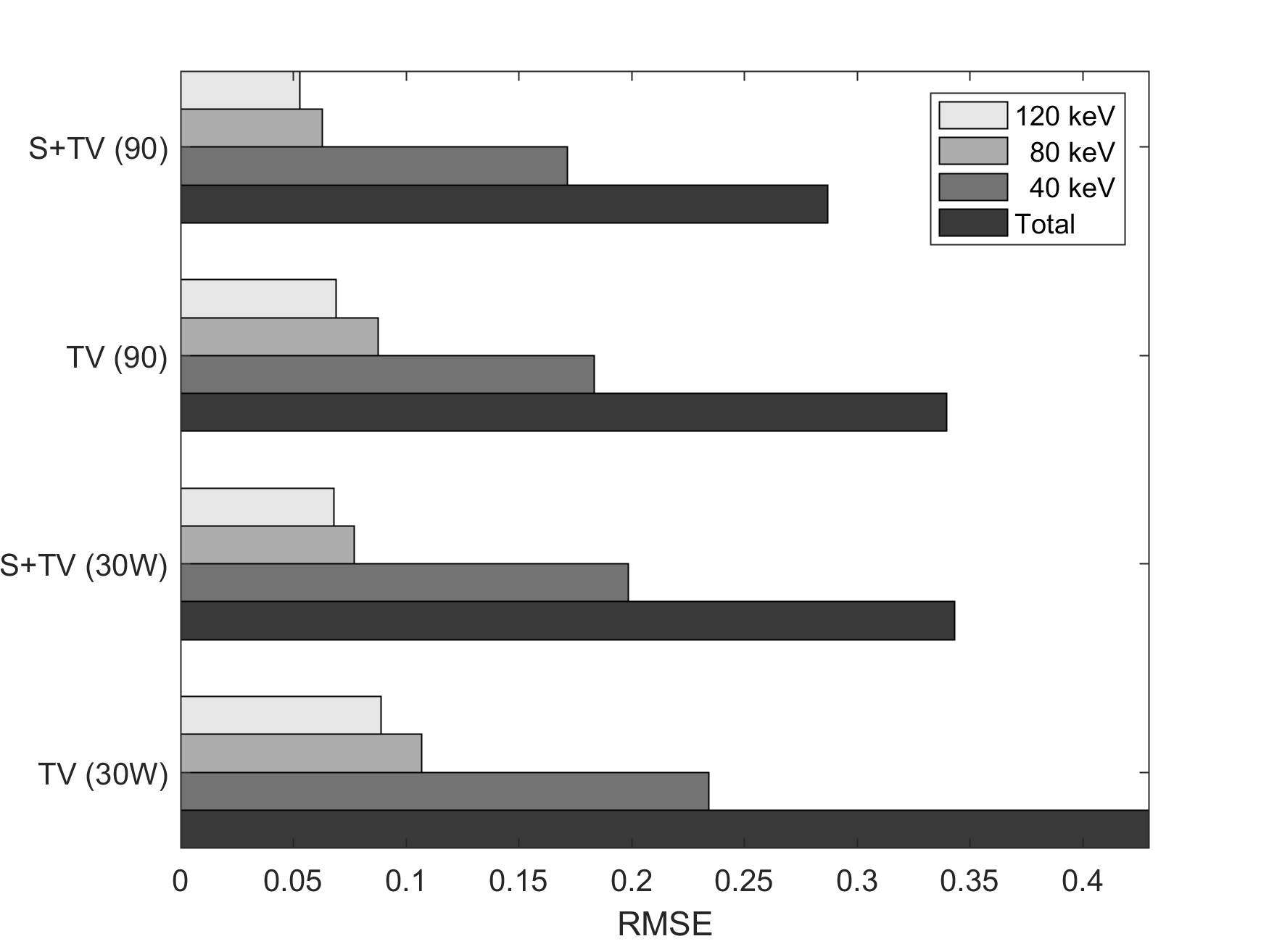}\\
  \caption{Error metrics for the TV and S+TV reconstructions shown in Figure \ref{fig:90vs30images}. Results labeled METHOD(90) are based on overall of $3\times 90 = 270$ projection images using the same 90 directions for each energy. Results labeled METHOD(30W) are based on overall of 90 projections using the non-overlapping 30+30+30 directions.}
  \label{fig:90vs30metrics}
\end{figure}

The results in Figures \ref{fig:SIM30} and \ref{fig:SIM30Metrics} reveal that the joint reconstruction models outperform the independent reconstruction models (No prior, TV) when using the same projection data of a total of 90 projection images. %Figure \ref{fig:90vs30} shows a comparison of TV and S+TV reconstructions using the same 
Figure \ref{fig:90vs30images} shows a comparison of TV and S+TV reconstructions using the same
measurement data of 90 projection images with the 30+30+30 direction non-overlapping geometry that was used in Fig. \ref{fig:SIM30} and also reconstructions using measurement data of 270 projection images using the same 90 directions for each energy. 
Error metrics for the reconstructions are shown in Figure \ref{fig:90vs30metrics}.
By comparison of the reconstructions S+TV(90) and TV(90), the S+TV outperforms the TV reconstructions also in this case when using classical overlapping projection geometry with the same 90 directions measured at each energy. More importantly, the comparison of TV(90) to S+TV(30W) reveals that they both perform approximately equally well even though S+TV(30W) uses only 90 projection images instead of the 270 used by TV(90). This finding suggests that the proposed combination of joint reconstruction and non-overlapping  projection direction geometry can facilitate significant reduction of the x-ray dose.

%\begin{figure}[p]
%  \centering
%  \includegraphics[width=0.45\textwidth]{SIM90ByCombining303030_RMSE_200207.png}\\
%  \includegraphics[width=0.75\textwidth]{SIM90ByCombining303030_181015_Zoom_T512.png}
%  \caption{Comparison of TV and S+TV using 90 projection directions. Results labeled METHOD(90) are based on overall of 
%$3\times 90 = 270$ projection images using the same 90 directions for each energy. Results labeled METHOD(30W) are based on overall of 90 projections using the non-overlapping 30+30+30 directions (the same geometry that is used in Figure \ref{fig:SIM30}).
%Top section: Error metrics. Bottom section: ROI of the reconstructed images.}
%  \label{fig:90vs30}
%\end{figure}

% -------------------------------------------------------------------
\subsection{Experimental Data}

\begin{figure}[p]
  \includegraphics[height=0.875\textheight]{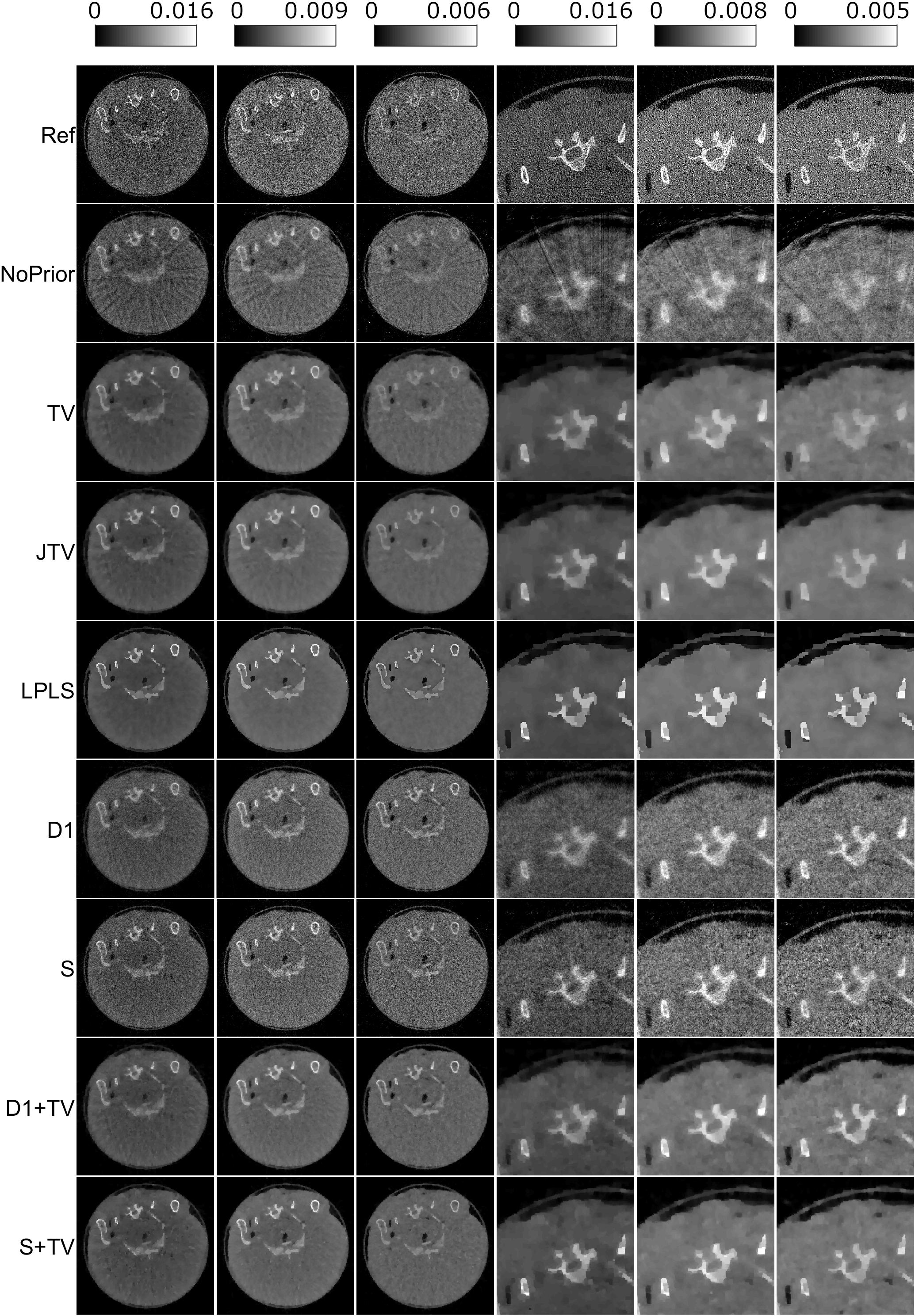}
  \caption{Reconstructed images for the experimental data test case.}
  \label{fig:EXP30}
\end{figure}

The reconstructed images for the experimental data test case are shown in Figure \ref{fig:EXP30}. The first three columns show the reconstructions for 
$E_1$, $E_2$, and $E_3$,
respectively, and the next three columns show a region of interest from the corresponding reconstructions. For reference, the first row (Ref) shows ground truth reconstructions computed with the least squares (No prior) model using the full 720 projection angle data for each energy.

The reconstructions of the experimental data test case support the findings of the simulation test case. The joint reconstructions (JTV, LPLS, D1, S, D1+TV, S+TV) are visually better than the reference single energy reconstructions (No prior, TV) and there is a difference in the amount of details between (TV, JTV, LPLS) and (D1, S), visible for example from the outer edge of the target in the zoomed in images. The combined regularization approaches (D1+TV, S+TV) provide again a good combination of the spatial details and noise level, noticeable for example in the visibility of the two small low value areas on the right hand side of the zoomed in images.

% -------------------------------------------------------------------
\section{Discussion and Conclusions}
\label{sec:Conclusions}
In this paper, we studied the performance of various structural similarity promoting joint reconstruction models using a novel low dose data acquisition protocol. The models were tested using both simulated and experimental measurement data and in both cases the linear parallel level sets (LPLS), joint total variation (JTV) and the combined structural similarity index based and total variation regularization (S+TV) performed the best. The regularization models that utilized only spectral regularization (S, D1) performed worse with respect to the RMSE and MSSIM fidelity metrics because they resulted in noisier reconstructions but nevertheless showed more details than the non-joint reconstruction approaches (No prior, TV). The combination of the joint reconstruction and the novel low dose projection protocol with non-overlapping projection directions was also compared to TV reconstructions using classical projection geometry, and was found to tolerate a clear reduction in the number of projection images, suggesting that the proposed approach can facilitate an effective way to reduce x-ray dose in multi-energy CT.

An implicit assumption in the proposed approach is that the target remains stationary during the imaging with different energies. However, with a data acquisition protocol that uses mutually non-overlapping projection angles and mid-scan energy switching, these motion issues should not, at least in theory, be any more problematic than with regular CT scanning. In addition, since the
proposed combination of joint reconstruction and non-overlapping projection directions can operate on a smaller number of projection images than the combination of conventional projection geometries and reconstruction methods, it could potentially also facilitate a shorter overall scanning time to mitigate the risks of motion artifacts.

\revmod{
In this study, we employed smoothed versions of the total variation and linear parallel level set regularization functionals and 
computed the optimization solutions using a Polak-Ribi\`ere non-linear conjugate gradient method. However, using the non-smooth versions of the regularization functionals can lead to sharper edges in the reconstructions and the optimization could be handled using well-known algorithms such as the Chambolle-Pock algorithm \cite{chambolle2011first, sidky2012convex} or the primal dual hybrid gradient algorithm \cite{zhu2008efficient}. The approach used here, on the other hand, has the benefit that the optimization using the smoothed functionals requires significantly fewer iterations (and thus less forward and back projection operations) than the minimization of the non-smooth functionals, implying significantly faster computation times, especially in large-scale reconstruction problems with three spatial dimensions or a temporal dimension, or both.}

\revmod{
The joint reconstruction approach used in this study can be applied directly to multi-energy CT data obtained with energy discriminating detectors. In such a case, the target would be imaged for each angle using a single exposure and the detector would distribute the measured photons to different energy bins, implying that each energy channel is measured at all projection angles. However, as indicated by comparison of S+TV(90) and TV(90) in Figures 
\ref{fig:90vs30images} and \ref{fig:90vs30metrics}, the proposed joint reconstruction approach would be beneficial also when working with data with overlapping projection angles.
}

\revmod{In this study, we employed sparse data which contained 30 projections at each energy, leading to an overall of 90 projection directions. In practical applications, the critical number of angles needed could be sought for by a similar simulated phantom based study that is proposed for limited angle tomography of the breast in \cite{sechopoulos2009optimization}.}

% ===================================================================
\section*{Acknowledgments}

This work was supported by the Academy of Finland (Project 312343, Finnish Centre of Excellence in Inverse Modelling and Imaging), the Jane and Aatos Erkko Foundation, and Business Finland project 6614/31/2016.

\bibliography{CTBib}

\begin{thebibliography}{10}

\bibitem{lell2015evolution}
Michael~M Lell, Joachim~E Wildberger, Hatem Alkadhi, John Damilakis, and Marc
  Kachelriess.
\newblock Evolution in computed tomography: The battle for speed and dose.
\newblock {\em Investigative Radiology}, 50(9):629--644, 2015.

\bibitem{dechiffre2014industrial}
L~De~Chiffre, S~Carmignato, J-P Kruth, R~Schmitt, and A~Weckenmann.
\newblock Industrial applications of computed tomography.
\newblock {\em CIRP Annals - Manufacturing Technology}, 64:655–677, 2014.

\bibitem{mccollough2012dose}
Cynthia~H McCollough, Guang~Hong Chen, Willi Kalender, Shuai Leng, Ehsan Samei,
  Katsuyuki Taguchi, Ge~Wang, Lifeng Yu, and Roderic~I Pettrigrew.
\newblock Achieving routine submillisievert {CT} scanning: report from the
  summit on management of radiation dose in {CT}.
\newblock {\em Radiology}, 264(2):567--580, 2012.

\bibitem{ginat2014technology}
Daniel~Thomas Ginat and Rajiv Gupta.
\newblock Advances in computed tomography imaging technology.
\newblock {\em Annual Review of Biomedical Engineering}, 16:431--453, 2014.

\bibitem{mccollough2015dual}
Cynthia~H McCollough, Shuai Leng, Lifeng Yu, and Joel~G Fletcher.
\newblock Dual-and multi-energy {CT}: principles, technical approaches, and
  clinical applications.
\newblock {\em Radiology}, 276(3):637--653, 2015.

\bibitem{goo2017dual}
Hyun~Woo Goo and Jin~Mo Goo.
\newblock Dual-energy {CT}: New horizon in medical imaging.
\newblock {\em Korean Journal of Radiology}, 18(4):555--569, 2017.

\bibitem{fornaro2011multi}
Juergen Fornaro, Sebastian Leschka, Dennis Hibbeln, Anthony Butler, Nigel
  Anderson, Gregor Pache, Hans Scheffel, Simon Wildermuth, Hatem Alkadhi, and
  Paul Stolzmann.
\newblock Dual- and multi-energy {CT}: approach to functional imaging.
\newblock {\em Insights into Imaging}, 2(2):149–--59, 2011.

\bibitem{forghani2018characterization}
R~Forghani and S~K Mukherji.
\newblock Advanced dual-energy {CT} applications for the evaluation of the soft
  tissues of the neck.
\newblock {\em Clinical Radiology}, 73(1):70--80, 2018.

\bibitem{marin2014characterization}
Daniele Marin, Daniel~T Boll, Achille Mileto, and Rendon~C Nelson.
\newblock State of the art: Dual-energy {CT} of the abdomen.
\newblock {\em Radiology}, 271(2):327--342, 2014.

\bibitem{nicolaou2012characterization}
Savvakis Nicolaou, Teresa Liang, Darra~T Murphy, Jeff~R Korzan, Hugue
  Ouellette, and Peter Munk.
\newblock Dual-energy {CT}: A promising new technique for assessment of the
  musculoskeletal system.
\newblock {\em American Journal of Roentgenology}, 199(5):S78--S86, 2012.

\bibitem{wong2018musculosceletal}
William~D Wong, Samad Shah, Nicholas Murray, Frances Walstra, Faisal Khosa, and
  Savvas Savvas~Nicolaou.
\newblock Advanced musculoskeletal applications of dual-energy computed
  tomography.
\newblock {\em Radiologic Clinics of North America}, 56(4):587--600, 2018.

\bibitem{yu2012mono}
Lifeng Yu, Shuai Leng, and Cynthia~H McCollough.
\newblock Dual-energy {CT}–based monochromatic imaging.
\newblock {\em American Journal of Roentgenology}, 199(5):S9--S15, 2012.

\bibitem{postma2012bone}
Alida~A Postma, Paul A~M Hofman, Annika A~R Stadler, Robert~J van Oostenbrugge,
  Maud P~M Tijssen, and Joachim~E Wildberger.
\newblock Dual-energy {CT} of the brain and intracranial vessels.
\newblock {\em American Journal of Roentgenology}, 199(5):S26--S33, 2012.

\bibitem{danad2015cardiac}
Ibrahim Danad, Zahi~A Fayad, Martin~J Willemink, and James~K Min.
\newblock New applications of cardiac computed tomography.
\newblock {\em JACC: Cardiovascular Interventions}, 8(6):710--723, 2015.

\bibitem{kalisz2017cardiac}
Kevin Kalisz, Sandra Halliburton, Suhny Abbara, Jonathon~A Leipsic, Moritz~H
  Albrecht, U~Joseph Schoepf, and Prabhakar Prabhakar~Rajiah.
\newblock Update on cardiovascular applications of multienergy {CT}.
\newblock {\em RadioGraphics}, 37(7):1955–--1974, 2017.

\bibitem{desantis2018cardiac}
Domenico De~Santis, Marwen Eid, Carlo~N De~Cecco, Brian~E Jacobs, Moritz~H
  Albrecht, Akos Varga-Szemes, Christian Tesche, Damiano Caruso, Andrea Laghi,
  and Uwe~Joseph Schoepf.
\newblock Dual-energy computed tomography in cardiothoracic vascular imaging.
\newblock {\em Radiologic Clinics of North America}, 56(4):521--534, 2018.

\bibitem{symons2017contrast}
Rolf Symons, Bernhard Krauss, Pooyan Sahbaee, Tyler Cork, Manu N.~Lakshmanan,
  David A.~Bluemke, and Amir Pourmorteza.
\newblock Photon-counting {CT} for simultaneous imaging of multiple contrast
  agents in the abdomen: An in vivo study.
\newblock {\em Medical Physics}, 44(10):5120--5127, 2017.

\bibitem{alvarez1976multienergy}
Robert~E Alvarez and Albert Macovski.
\newblock Energy-selective reconstructions in {X}-ray computerised tomography.
\newblock {\em Physics in Medicine \& Biology}, 21(5):733--744, 1976.

\bibitem{kalender2014dose}
Willi~A Kalender.
\newblock Dose in {X}-ray computed tomography.
\newblock {\em Physics in Medicine \& Biology}, 59(3):R129--R150, 2014.

\bibitem{tubiana2009radbio}
Maurice Tubiana, Ludwig~E Feinendegen, Chichuan Yang, and Joseph~M Kaminski.
\newblock The linear no-threshold relationship is inconsistent with radiation
  biologic and experimental data.
\newblock {\em Radiology}, 251(1):13--22, 2009.

\bibitem{hendee2012risks}
William~R Hendee and Michael~K O’Connor.
\newblock Radiation risks of medical imaging: Separating fact from fantasy.
\newblock {\em Radiology}, 264(2):312--321, 2012.

\bibitem{icrp103}
International~Commission on~Radiological~Protection.
\newblock {ICRP} publication 103: The 2007 recommendations of the international
  commission on radiological protection.
\newblock {\em Annals of the ICRP}, 37(2--4), 2007.

\bibitem{icrp105}
International~Commission on~Radiological~Protection.
\newblock {ICRP} publication 105: Radiological protection in medicine.
\newblock {\em Annals of the ICRP}, 37(6), 2007.

\bibitem{buzug2008ct}
Thorsten~M Buzug.
\newblock {\em Computed Tomography: From Photon Statistics to Modern Cone-Beam
  {CT}}.
\newblock Springer, 2008.

\bibitem{pan2009fbp}
Xiaochuan Pan, Emil~Y Sidky, and Michael Vannier.
\newblock Why do commercial {CT} scanners still employ traditional, filtered
  back-projection for image reconstruction?
\newblock {\em Inverse Problems}, 25(12):123009, 2009.

\bibitem{hamalainen2013sparse}
Keijo Hamalainen, Aki Kallonen, Ville Kolehmainen, Matti Lassas, Kati
  Niinimaki, and Samuli Siltanen.
\newblock Sparse tomography.
\newblock {\em SIAM Journal on Scientific Computing}, 35(3):B644--B665, 2013.

\bibitem{sechopoulos2009optimization}
Ioannis Sechopoulos and Caterina Ghetti.
\newblock Optimization of the acquisition geometry in digital tomosynthesis of
  the breast.
\newblock {\em Medical physics}, 36(4):1199--1207, 2009.

\bibitem{pan2014structural}
Adam Pan, Ling Xu, Justin Lee, Rajiv Gupta, and George Barbastathis.
\newblock Structural similarity regularization of {X}-ray transport of
  intensity phase retrieval.
\newblock In {\em Computational Optical Sensing and Imaging}, pages CW2C--4.
  Optical Society of America, 2014.

\bibitem{yang2012structural}
Shuyuan Yang, Yaxin Sun, Yiguang Chen, and Licheng Jiao.
\newblock Structural similarity regularized and sparse coding based
  super-resolution for medical images.
\newblock {\em Biomedical Signal Processing and Control}, 7(6):579--590, 2012.

\bibitem{rigie2015vtv}
David~S Rigie and Patrick~J La~Rivi{\'e}re.
\newblock Joint reconstruction of multi-channel, spectral {CT} data via
  constrained total nuclear variation minimization.
\newblock {\em Physics in Medicine \& Biology}, 60(5):1741–1762, 2015.

\bibitem{rigie2017vtv}
David~S Rigie, Adrian~A Sanchez, and Patrick~J La~Rivi{\'e}re.
\newblock Assessment of vectorial total variation penalties on realistic
  dual-energy {CT} data.
\newblock {\em Physics in Medicine \& Biology}, 62(8):3284–3298, 2017.

\bibitem{kim2015patch}
Kyungsang Kim, Jong~Chul Ye, William Worstell, Jinsong Ouyang, Yothin
  Rakvongthai, Georges El~Fakhri, and Quanzheng Li.
\newblock Sparse-view spectral {CT} reconstruction using spectral patch-based
  low-rank penalty.
\newblock {\em IEEE Transactions on Medical Imaging}, 34(3):748–760, 2015.

\bibitem{zhang2017dictionary}
Yanbo Zhang, Xuanqin Mou, Ge~Wang, and Hengyong Yu.
\newblock Tensor-based dictionary learning for spectral {CT} reconstruction.
\newblock {\em IEEE Transactions on Medical Imaging}, 36(1):142--154, 2017.

\bibitem{wu2018dictionary}
Weiwen Wu, Yanbo Zhang, Qian Wang, Fenglin Liu, Peijun Chen, and Hengyong Yu.
\newblock Low-dose spectral {CT} reconstruction using image gradient
  $\ell$0–norm and tensor dictionary.
\newblock {\em Applied Mathematical Modelling}, 63:538--557, 2018.

\bibitem{kazantsev2018joint}
Daniil Kazantsev, Jakob~S J{\o}rgensen, Martin~S Andersen, William~RB
  Lionheart, Peter~D Lee, and Philip~J Withers.
\newblock Joint image reconstruction method with correlative multi-channel
  prior for x-ray spectral computed tomography.
\newblock {\em Inverse Problems}, 34(6):064001, 2018.

\bibitem{gao2011prism}
Hao Gao, Hengyong Yu, Stanley Osher, and Ge~Wang.
\newblock Multi-energy {CT} based on a prior rank, intensity and sparsity model
  ({PRISM}).
\newblock {\em Inverse Problems}, 27(11):115012, 2011.

\bibitem{yang2017prism}
Q~Yang, W~Cong, and G~Wang.
\newblock Superiorization-based multi-energy {CT} image reconstruction.
\newblock {\em Inverse Problems}, 33(4):044014, 2017.

\bibitem{semerci2014tensor}
Oguz Semerci, Ning Hao, Misha~E Kilmer, and Eric~L Miller.
\newblock Tensor-based formulation and nuclear norm regularization for
  multienergy computed tomography.
\newblock {\em IEEE Transactions on Image Processing}, 23(4):1678--1693, 2014.

\bibitem{niu2018nonlocal}
Shanzhou Niu, Gaohang Yu, Jianhua Ma, and Jing Wang.
\newblock Nonlocal low-rank and sparse matrix decomposition for spectral {CT}
  reconstruction.
\newblock {\em Inverse Problems}, 34(2):024003, 2018.

\bibitem{chen2017nonconvex}
Buxin Chen, Zheng Zhang, Emil~Y Sidky, Dan Xia, and Xiaochuan Pan.
\newblock Image reconstruction and scan configurations enabled by
  optimization-based algorithms in multispectral {CT}.
\newblock {\em Physics in Medicine \& Biology}, 62(22):8763–--8793, 2017.

\bibitem{blomgren1998color}
Peter Blomgren and Tony~F Chan.
\newblock Color {TV}: total variation methods for restoration of vector-valued
  images.
\newblock {\em IEEE transactions on image processing}, 7(3):304--309, 1998.

\bibitem{ehrhardt2014vector}
Matthias~Joachim Ehrhardt and Simon~R Arridge.
\newblock Vector-valued image processing by parallel level sets.
\newblock {\em IEEE Transactions on Image Processing}, 23(1):9--18, 2014.

\bibitem{rasch2018dynamic}
Julian Rasch, Ville Kolehmainen, Riikka Nivaj{\"a}rvi, Mikko Kettunen, Olli
  Gr{\"o}hn, Martin Burger, and Eva-Maria Brinkmann.
\newblock Dynamic {MRI} reconstruction from undersampled data with an
  anatomical prescan.
\newblock {\em Inverse Problems}, 34(7):074001, 2018.

\bibitem{gursoy2015hyperspectral}
Doga G{\"u}rsoy, Tekin Bi{\c{c}}er, Antonio Lanzirotti, Matthew~G Newville,
  and Francesco De~Carlo.
\newblock Hyperspectral image reconstruction for {X}-ray fluorescence
  tomography.
\newblock {\em Optics express}, 23(7):9014--9023, 2015.

\bibitem{wang2004image}
Zhou Wang, Alan~C Bovik, Hamid~R Sheikh, and Eero~P Simoncelli.
\newblock Image quality assessment: from error visibility to structural
  similarity.
\newblock {\em IEEE transactions on image processing}, 13(4):600--612, 2004.

\bibitem{michielsen2018dose}
Koen Michielsen, Christian Fedon, James Nagy, and Ioannis Sechopoulos.
\newblock Dose reduction in breast {CT} by spectrum switching.
\newblock In {\em 14th International Workshop on Breast Imaging (IWBI 2018)},
  volume 10718, page 107180J. International Society for Optics and Photonics,
  2018.

\bibitem{osullivan2007alternating}
Joseph~A O'Sullivan and Jasenka Benac.
\newblock Alternating minimization algorithms for transmission tomography.
\newblock {\em IEEE Transactions on Medical Imaging}, 26(3):283--297, 2007.

\bibitem{chung2010numerical}
Julianne Chung, James~G Nagy, and Ioannis Sechopoulos.
\newblock Numerical algorithms for polyenergetic digital breast tomosynthesis
  reconstruction.
\newblock {\em SIAM Journal on Imaging Sciences}, 3(1):133--152, 2010.

\bibitem{elbakri2002segmentation}
Idris~A Elbakri and Jeffrey~A Fessler.
\newblock Segmentation-free statistical image reconstruction for polyenergetic
  {X}-ray computed tomography.
\newblock In {\em Proceedings IEEE International Symposium on Biomedical
  Imaging}, pages 828--831. IEEE, 2002.

\bibitem{mccketty1998energy}
Marlene~H. McKetty.
\newblock The {AAPM/RSNA} physics tutorial for residents.
\newblock {\em Radiographics}, 18(1):151--163, 1998.

\bibitem{nist2004xray}
National~Institute of~Standards and Technology Physical~Measurement Laboratory.
\newblock {X}-ray mass attenuation coefficients: {NIST} standard reference
  database 126, 2004.

\bibitem{poludniowski2009spekcalc}
G~Poludniowski, G~Landry, F~DeBlois, P~M Evans, and F~Verhaegen.
\newblock {SpekCalc}: a program to calculate photon spectra from tungsten anode
  x-ray tubes.
\newblock {\em Physics in Medicine and Biology}, 54:N433–--N438, 2009.

\bibitem{natterer1986mathematics}
Frank Natterer.
\newblock {\em The mathematics of computerized tomography}, volume~32.
\newblock Siam, 1986.

\bibitem{donoho2006compressed}
David~L Donoho.
\newblock Compressed sensing.
\newblock {\em IEEE Transactions on Information Theory}, 52(4):1289--1306,
  2006.

\bibitem{bubba2017shearlet}
TA~Bubba, M~M{\"a}rz, Z~Purisha, M~Lassas, and S~Siltanen.
\newblock Shearlet-based regularization in sparse dynamic tomography.
\newblock In {\em Wavelets and Sparsity XVII}, volume 10394, page 103940Y.
  International Society for Optics and Photonics, 2017.

\bibitem{siltanen2003statistical}
Samuli Siltanen, Ville Kolehmainen, Seppo J{\"a}rvenp{\"a}{\"a}, JP~Kaipio,
  P~Koistinen, M~Lassas, J~Pirttil{\"a}, and E~Somersalo.
\newblock Statistical inversion for medical x-ray tomography with few
  radiographs: {I.} general theory.
\newblock {\em Physics in Medicine \& Biology}, 48(10):1437, 2003.

\bibitem{kolehmainen2003statistical}
Ville Kolehmainen, Samuli Siltanen, Seppo J{\"a}rvenp{\"a}{\"a}, Jari~P Kaipio,
  P~Koistinen, M~Lassas, J~Pirttil{\"a}, and E~Somersalo.
\newblock Statistical inversion for medical x-ray tomography with few
  radiographs: {II.} application to dental radiology.
\newblock {\em Physics in Medicine \& Biology}, 48(10):1465, 2003.

\bibitem{sidky2008image}
Emil~Y Sidky and Xiaochuan Pan.
\newblock Image reconstruction in circular cone-beam computed tomography by
  constrained, total-variation minimization.
\newblock {\em Physics in Medicine \& Biology}, 53(17):4777, 2008.

\bibitem{hamalainen2014total}
Keijo H{\"a}m{\"a}l{\"a}inen, Lauri Harhanen, Andreas Hauptmann, Aki Kallonen,
  Esa Niemi, and Samuli Siltanen.
\newblock Total variation regularization for large-scale {X}-ray tomography.
\newblock {\em International Journal of Tomography \& Simulation}, 25(1):1--25,
  2014.

\bibitem{rudin1992nonlinear}
Leonid~I Rudin, Stanley Osher, and Emad Fatemi.
\newblock Nonlinear total variation based noise removal algorithms.
\newblock {\em Physica D: nonlinear phenomena}, 60(1-4):259--268, 1992.

\bibitem{ehrhardt2014joint}
Matthias~J Ehrhardt, Kris Thielemans, Luis Pizarro, David Atkinson,
  S{\'e}bastien Ourselin, Brian~F Hutton, and Simon~R Arridge.
\newblock Joint reconstruction of {PET-MRI} by exploiting structural
  similarity.
\newblock {\em Inverse Problems}, 31(1):015001, 2014.

\bibitem{ehrhardt2016pet}
Matthias~J Ehrhardt, Pawel Markiewicz, Maria Liljeroth, Anna Barnes, Ville
  Kolehmainen, John~S Duncan, Luis Pizarro, David Atkinson, Brian~F Hutton,
  S{\'e}bastien Ourselin, et~al.
\newblock {PET} reconstruction with an anatomical {MRI} prior using parallel
  level sets.
\newblock {\em IEEE transactions on medical imaging}, 35(9):2189--2199, 2016.

\bibitem{kolehmainen2019incorporating}
Ville Kolehmainen, Matthias~J Ehrhardt, and Simon~R Arridge.
\newblock Incorporating structural prior information and sparsity into {EIT}
  using parallel level sets.
\newblock {\em Inverse Problems \& Imaging}, 13(2):285--307, 2019.

\bibitem{wang2009mean}
Zhou Wang and Alan~C Bovik.
\newblock Mean squared error: Love it or leave it? a new look at signal
  fidelity measures.
\newblock {\em IEEE signal processing magazine}, 26(1):98--117, 2009.

\bibitem{segars2008xcat}
W~Paul Segars, Mahadevappa Mahesh, Thomas~J Beck, Eric~C Frey, and Benjamin M~W
  Tsui.
\newblock Realistic {CT} simulation using the {4D} {XCAT} phantom.
\newblock {\em Medical Physics}, 35(8):3800--3808, 2008.

\bibitem{segars2010xcat}
W~Paul Segars, Gregory~M Sturgeon, S~Mendonca, Jason Grimes, and Benjamin M~W
  Tsui.
\newblock {4D} {XCAT} phantom for multimodality imaging research.
\newblock {\em Medical Physics}, 37(9):4902--4915, 2010.

\bibitem{van2016fast}
Wim van Aarle, Willem~Jan Palenstijn, Jeroen Cant, Eline Janssens, Folkert
  Bleichrodt, Andrei Dabravolski, Jan De~Beenhouwer, K~Joost Batenburg, and Jan
  Sijbers.
\newblock Fast and flexible {X}-ray tomography using the {ASTRA} toolbox.
\newblock {\em Optics express}, 24(22):25129--25147, 2016.

\bibitem{van2015astra}
Wim van Aarle, Willem~Jan Palenstijn, Jan De~Beenhouwer, Thomas Altantzis, Sara
  Bals, K~Joost Batenburg, and Jan Sijbers.
\newblock The {ASTRA} toolbox: A platform for advanced algorithm development in
  electron tomography.
\newblock {\em Ultramicroscopy}, 157:35--47, 2015.

\bibitem{astrawebpages}
{The ASTRA Toolbox}.
\newblock \url{http://www.astra-toolbox.com/}.

\bibitem{polak1969note}
Elijah Polak and Gerard Ribiere.
\newblock Note sur la convergence de m{\'e}thodes de directions conjugu{\'e}es.
\newblock {\em Revue fran{\c{c}}aise d'informatique et de recherche
  op{\'e}rationnelle. S{\'e}rie rouge}, 3(16):35--43, 1969.

\bibitem{chambolle2011first}
Antonin Chambolle and Thomas Pock.
\newblock A first-order primal-dual algorithm for convex problems with
  applications to imaging.
\newblock {\em Journal of mathematical imaging and vision}, 40(1):120--145,
  2011.

\bibitem{sidky2012convex}
Emil~Y Sidky, Jakob~H J{\o}rgensen, and Xiaochuan Pan.
\newblock Convex optimization problem prototyping for image reconstruction in
  computed tomography with the chambolle--pock algorithm.
\newblock {\em Physics in Medicine \& Biology}, 57(10):3065, 2012.

\bibitem{zhu2008efficient}
Mingqiang Zhu and Tony Chan.
\newblock An efficient primal-dual hybrid gradient algorithm for total
  variation image restoration.
\newblock {\em UCLA CAM Report}, 34, 2008.

\end{thebibliography}

\end{document}